\documentclass[aps,prd,preprint,showpacs,nofootinbib,preprintnumbers]{revtex4}

\usepackage{epsfig}
\usepackage{amsmath}
\usepackage{amssymb}
\usepackage[dvipdfm]{hyperref}
\def\e{{\,\rm e}\,}
\def\d{{\rm d}}

\def\D{{\cal D}}

\newcommand{\rf}[1]{(\ref{#1})}
\newcommand{\eq}[1]{Eq.~(\ref{#1})}
\def\be{\begin{equation}}
\def\ee{\end{equation}}
\def\beq{\begin{equation}}
\def\eeq{\end{equation}}
\def\bea{\begin{eqnarray}}
\def\eea{\end{eqnarray}}
\def\LA{\left\langle}
\def\RA{\right\rangle}
\newcommand{\non}{\nonumber \\*}

\begin{document}

\preprint{ITEP--TH--55/09}

\title{Path Integral over Reparametrizations:\\
L\'evy Flights versus Random Walks}

\author{Pavel Buividovich,%
\footnote{On leave from JIPNR ``Sosny'',
Nat.\ Acad.\ of Science, Acad.\ Krasin str.\ 99, 220109 Minsk, Belarus}
Yuri Makeenko%
\footnote{Also at the Institute for Advanced Cycling,
Blegdamsvej 19, 2100 Copenhagen \O, Denmark}}

\affiliation{Institute of Theoretical and Experimental Physics\\
B.~Cheremushkinskaya 25, 117218 Moscow, Russia}
\email{buividovich@itep.ru,  makeenko@itep.ru}

\date{February 11, 2010}

\begin{abstract}
 We investigate the properties of the path integral over reparametrizations (or the boundary value of the Liouville field in string theory). Discretizing the path integral, we apply the Metropolis--Hastings algorithm to numerical simulations of a proper (subordinator) stochastic process and find that typical trajectories are not Brownian but rather have discontinuities of the type of L\'evy's flights. We study a fractal structure of these trajectories and show that their Hausdorff dimension is zero. We confirm thereby previous results on QCD scattering amplitudes by analytical and numerical calculations. We also perform Monte-Carlo simulations of the path integral over reparametrization in the effective string ansatz for a circular Wilson loop and discuss their subtleties associated with the discretization of Douglas' functional.
\end{abstract}


\maketitle

\section{Introduction}

 Path integrals in Quantum Field Theory are usually calculated with the Gaussian Wiener measure, which leads to continuous trajectories because of the presence of $\int \dot x{}^2$ in the action. Typical trajectories are then of the Brownian type, like those for the standard random walks.

 Another kind of the measure is associated  with the path integral over reparametrizations or diffeomorphisms (i.e.\ functions with a nonnegative derivative) of a curve, which emerge in many interesting physical and mathematical problems,
e.g.\ the solution~\cite{Dou31} of the problem of Plateau. An important example is an integration over the boundary value of the Liouville field in Polyakov's formulation of string theory~\cite{Pol81,Alv83,DOP84,CMNP85}. Usually, this path integral decouples~\cite{Pol87,ADN85} in the critical number of dimensions for tachyonic scalars or massless vectors, but is present generically for noncritical string.

 More recently the path integral over reparametrizations has appeared in an effective string ansatz~\cite{Pol97} (see also Ref.~\cite{Ryc02}) for the Wilson loops $W(C)$ in large-$N$ QCD and was crucial in deriving QCD scattering amplitudes~\cite{MO08,MO09} from these Wilson loops. The measure for the integration over reparametrizations was defined, roughly speaking, as
\begin{equation}
\int_{s(\tau_0)=s_0}^{s(\tau_f)=s_f} \D_{\rm diff} s \cdots =
\int \prod_{\tau=\tau_0}^{\tau_f} \frac{\d s(\tau)}{s'(\tau)}\cdots \qquad
s'(\tau)\geq 0\,,
\label{defdiffmea}
\end{equation}
where $\tau$ is a parameter, parametrizing the loop $C$, e.g.\ the proper length. Splitting $[\tau_0, \tau_f]$ into $N$ equidistant intervals, a discretization of the measure~\rf{defdiffmea} can be written as
\begin{equation}
\int_{s_0}^{s_f}\D_{\rm diff} s \cdots =\lim_{N \to \infty}
\prod_{i=2}^{N-1}\int_{s_0}^{s_{i+1}}
 \frac{\d s_i}{(s_{i+1}-s_{i})}\int_{s_0}^{s_{2}}
 \frac{\d s_1}{(s_{2}-s_{1})(s_{1}-s_{0})}\cdots \,,
\label{discdefdiffmea}
\end{equation}
where the integral goes over $(N-1)$ subordinated values $s_0\leq \cdots \leq s_{i-1} \leq s_i \leq \cdots \leq s_N=s_f$. Discretizing the continuum formula $s'=\exp[-\varphi]$ and relating reparametrizations to the boundary value of the Liouville field $\varphi(\tau)$%
\footnote{Our definition of $\varphi(\tau)$ is as in Ref.~\cite{Alv83} and differs from the standard one in string theory by a factor of $2$.}
by $s_i-s_{i-1}=\exp[-\varphi_i]$, we can rewrite \eq{discdefdiffmea} as
\begin{equation}
\int_{s_0}^{s_f}\D_{\rm diff} s \cdots =
\lim_{N \to \infty}
\prod_{i=1}^{N} \int_{-\infty}^{+\infty} {\d \varphi_i}\,
\delta^{(1)}\Big( s_f-s_0-\sum_{j=1}^{N} \e^{-\varphi_j} \Big)\cdots ,
\label{vvvi}
\end{equation}
where the only restriction on the values of $\varphi_i$'s is given by the one-dimensional delta-function involved.

 It is clear from the representation~\rf{vvvi} (or \rf{discdefdiffmea}) that there is no factor in the measure which makes typical trajectories to be continuous, as distinct from the Wiener measure. Therefore, a question arises as to what type of trajectories is essential in the path integral over reparametrizations, the answer to which will depend, in general, on the form of the integrand.

 In the present paper we investigate the properties of the path integral over reparametrizations by applying analytical and numerical methods to its discretized version \rf{discdefdiffmea}. We reformulate the problem in the language of a proper (subordinator) stochastic process and find that typical trajectories are in general not Brownian but rather may have discontinuities of the type of L\'evy's flights. We study a fractal structure of these trajectories and show that their Hausdorff dimension varies between zero and one, depending on the form of the integrand.
In particular, our analytical and numerical results confirm the discretization and heuristic consideration of the path integral over reparametrizations in QCD scattering amplitudes of Ref.~\cite{MO09}

 In Sect.~\ref{s:2} we formulate the path integral over reparametrizations as a subordinated stochastic process and introduce two types of regularizations of the measure: by the gamma-subordinator, which is convenient for exact analytical studies, and the logarithmic subordinator, which is more convenient for numerical simulations. We demonstrate that sample trajectories for the case when the integrand in \eq{discdefdiffmea} equals one, as is the case for the off-shell QCD scattering amplitudes~\cite{MO09}, are expected to have zero Hausdorff dimension.

 In Sect.~\ref{s:3} we apply the Metropolis--Hastings algorithm to numerical simulations of a proper (subordinator) stochastic process and find that typical trajectories  have discontinuities known as L\'evy's flights. Exact analytical results for the gamma-subordinator as well as numerical results for the logarithmic subordinator, defined in Sect.~\ref{s:2}, indicate that sample trajectories have a fractal structure with the Hausdorff dimension equal to zero.

 In Sect.~\ref{s:4} we numerically investigate the asymptotic effective string ansatz~\cite{Pol97} for the Wilson loops in large-$N$ QCD, which involves the path integral over reparametrizations, for a circle. Because of the presence of the Douglas integral~\cite{Dou31} in the action, the model reminds ones \cite{AYH70,Kos76,AES82, HZ97,LZ03,Zam06} of one-dimensional statistical mechanics, while the difference resides in the measure. The expectation that the Hausdorff dimension of sample trajectories decreases from one at large radii of the circle to zero at small radii is supported by our numerical results. There are, however, subtleties of implementing numerical calculations with a step-wise discretization of the Douglas integral that does not properly suppress trajectories with large discontinuities. We resolve this problem by using a polygon-wise discretization.

\section{Integration over reparametrizations\label{s:2}}

 The measure in the path integral over reparametrizations (or diffeomorphisms) --- the functions $s(t)$ with non-negative derivatives $s'(t)\geq 0$ --- is defined as \cite{MO09}
\bea
\D_{\rm diff} s&=&\lim_{N\to\infty} \D_{\rm diff}^{(N)} s \non
 \D_{\rm diff}^{(N)} s &=&
P\left(s_N-s_{N-1}\right)\prod_{i=1}^{N-1}
 P\left(s_i-s_{i-1}\right), \qquad
s_N\geq \cdots \geq s_i \geq s_{i-1}\geq \cdots \geq s_0\,,
\label{measure}
\eea
where $s_0$ and $s_N\equiv s_f$ are fixed initial and final points:
\be
s(s_0)=s_0, \quad s(s_f)=s_f\,.
\label{boundaryconds}
\ee
Sequences $s_{i}$ with $s_{i+1} > s_{i}$ are called {\em subordinators} in the theory of stochastic processes, and the sequences with fixed boundary conditions are known as {\em subordinators that hit a boundary}.\cite{App04}

 The distribution $P\left(\Delta s_i \right) $ was chosen in \cite{MO09} as
\be
P\left(\Delta s_i \right) =\frac{1}{\Gamma(\delta_i)
\left(\Delta s_i \right)^{1-\delta_i}}
\label{gamma-s}
\ee
with all equal $\delta_i=\delta>0$. It reminds the probability density of the gamma-subordinator%
\footnote{This probability density~\cite{App04} (displayed in \eq{36} below) involves the additional exponential $\e^{-b\Delta s_i}$, which cancels at the intermediate points. But it is important to provide the normalizability and an IR cutoff for the associated random walk problem. This factor is not essential for $b(s_f-s_0)\ll 1$.}.  The measure for integrating over reparametrizations is approached when $N\to\infty$ with $N \delta \to 0$, while $N$ has the meaning of the number of steps in the stochastic process.

We shall also consider another discretization of the measure with
\be
P\left(\Delta s_i \right) =\frac{1}{\ln(1/\varepsilon_i)
\left(\Delta s_i +\varepsilon_i \right)},
\label{log-s}
\ee
where $\varepsilon_i$'s are all the same ($\varepsilon_i=\varepsilon$).
In an analogy with the gamma-subordinator, we shall call the process caused by \eq{log-s} as the {\em logarithmic subordinator}. To describe the desired continuum limit, we have to take the limit $N\to\infty$ with $N\varepsilon\to 0$ as will be explained below.

The consecutive $N\!-\!1$-fold integral with the measure \rf{gamma-s} can be easily evaluated by using the formula
\be
\int_{s_{i-1}}^{s_{i+1}} \frac {\d s_i}{
\left(s_{i+1}- s_i \right)^{1-\delta_{i+1}}
\left(s_{i}- s_{i-1} \right)^{1-\delta_i}}
=\frac{\Gamma(\delta_{i})\Gamma(\delta_{i+1})}
{\Gamma(\delta_{i}+\delta_{i+1})}
\frac1{\left(s_{i+1}- s_{i-1} \right)^{1-\delta_{i}-\delta_{i+1}}}.
\label{main}
\ee
We find
\be
\int_{s_0}^{s_N=s_f}\D_{\rm diff}^{(N)} s = \frac{1}{\Gamma(N \delta)}
\frac{1}{\left(s_{N}- s_{0} \right)^{1-N\delta}},
\label{propaN}
\ee
which is an exact formula for arbitrary $N$ and $\delta>0$. In the functional limit of $N\to\infty$ with $N\delta\to0$ we get from \eq{propaN}
\be
\int_{s_0}^{s_f}\D_{\rm diff}^{(N)} s  \stackrel{N\delta\ll1}
\longrightarrow N \delta
\frac{1}{\left(s_{f}- s_{0} \right)^{1-N\delta}}
\label{propa}
\ee
recovering for $(s_f-s_0)\sim 1$ Eq.~(D8) of \cite{MO09}.

The structure of \eq{main} prompts to interpret $\delta$ as a time step of the associated stochastic process, while $\tau=N\delta$ is the corresponding time variable. Then
\be
\d s_f \int_{s_0}^{s_f}\D_{\rm diff}^{(N)} s
\label{proba}
\ee
describes the probability to ``propagate'' from $s_0$ to the interval $\left[s_f,s_f+\d s_f \right]$ during the time $\tau=N\delta$. For small $\tau$ and $(s_f-s_0)$ we can introduce the scaling variable
\be
z=\tau \ln \frac1{(s_f-s_0)} \,,
\ee
which is analogous to the usual scaling variable $(s_f-s_0)^2/\tau$ for the Gaussian random walks, and rewrite the probability~\rf{proba} as
\be
\frac{ \tau \d s_f}{\left(s_{f}- s_{0} \right)^{1-\tau}}
=\d z \e^{-z} \,,
\ee
which manifestly demonstrates scaling with
\be
(s_f-s_0)\sim \e^{-1/\tau}\,.
\label{scaling}
\ee
This supersedes the well-know relation $(s_f-s_0)^2\sim\tau$ for the Brownian motion (whose Hausdorff dimension equals two). We conclude, therefore, that for small $\tau$ and $(s_f-s_0)$ the Hausdorff dimension, which characterizes fractal properties of the given stochastic process, is zero. In Sect.~\ref{ss:3b} we shall extend this result to $(s_f-s_0)\sim1$.

We can also define the averages with respect to the measure~\rf{measure}, a most important of which is of the type
\be
\LA \left(\frac{(s_{K+1}-s_K)(s_{K}-s_{K-1})}{(s_{K+1}-s_{K-1})}\right)^a \RA
\equiv \frac{\int_{s_0}^{s_f}\D_{\rm diff}^{(N)} s
\left(\frac{(s_{K+1}-s_K)(s_{K}-s_{K-1})}{(s_{K+1}-s_{K-1})}\right)^a }
{\int_{s_0}^{s_f}\D_{\rm diff}^{(N)} s }\,.
\label{avera}
\ee
It appears in the calculation of QCD scattering amplitudes using the principal value prescription of \cite{MO09}
when $p(t(s))$ has a discontinuity $\Delta p_K$ at $s=s_K$ ($1\ll K \ll N$) but is constant for $s_0<s<s_K$ and $s_K<s<s_N=s_f$, so that $a=\alpha' \Delta p_K^2>0$ for Euclidean momenta. Using repeatedly \eq{main}, we obtain
\be
\LA \left(\frac{(s_{K+1}-s_K)(s_{K}-s_{K-1})}{(s_{K+1}-s_{K-1})}\right)^a \RA
=\frac{\Gamma(N\delta)}{\Gamma(a+N\delta)}
\frac{\Gamma^2(a+\delta)}{\Gamma^2(\delta)}
\frac{\Gamma(a+2\delta)}{\Gamma(2a+2\delta)} (s_N-s_0)^a.
\label{compliN}
\ee
This formula is again exact and compliments the consideration in Appendix~D of \cite{MO09} because
\bea
\protect{\rf{compliN}}&\stackrel{N\delta\ll1}\longrightarrow& \frac \delta N
\frac{\Gamma^2(a)}{\Gamma(2a)} (s_N-s_0)^a \non &=&
\frac\delta N \,(s_f-s_0)
\int_{s_0}^{s_f}  \d s_K
\frac1{(s_{f}-s_K)}
\left(\frac{(s_{f}-s_K)(s_{K}-s_{0})}{(s_{f}-s_{0})}\right)^a
\frac1{(s_{K}-s_{0})}
\label{compli}
\eea
as is given by Eq.~(D13) of \cite{MO09}.

Formulas, which are pretty much similar to Eqs.~\rf{propa} and \rf{compli}, can be derived also for the measure, given by Eqs.~\rf{measure}, \rf{log-s}, in the logarithmic approximation starting from the integral
\be
\int_{s_{i-1}}^{s_{i+1}} \frac {\d s_i}{
\left(s_{i+1}- s_i +\varepsilon_{i+1}\right)
\left(s_{i}- s_{i-1}+\varepsilon_i \right)}=
\frac{\ln\frac{(s_{i+1}-s_{i-1})}{\varepsilon_{i+1}}
+\ln\frac{(s_{i+1}-s_{i-1})}{\varepsilon_{i}}}
{\left(s_{i+1}- s_{i-1}+\varepsilon_i +\varepsilon_{i+1}\right)}.
\ee
Subsequent integrations can be performed with the logarithmic accuracy using the formula
\bea
\lefteqn{\int_{s_0}^{s_f} \d s
\frac{\ln^{N_2-1}\frac{s_f-s}{\varepsilon_2}
\ln^{N_1-1}\frac{s-s_0}{\varepsilon_1}}
{\left(s_{f}- s +\varepsilon_{2}\right)
\left(s- s_{0}+\varepsilon_1 \right)}} \non &=&
\frac{\frac{1}{N_2}\ln^{N_2}\frac{(s_{f}-s_{0})}{\varepsilon_{2}}
\ln^{N_1-1}\frac{(s_{f}-s_{0})}{\varepsilon_{1}}
+\frac{1}{N_1}\ln^{N_2-1}\frac{(s_{f}-s_{0})}{\varepsilon_{2}}
\ln^{N_1}\frac{(s_{f}-s_{0})}{\varepsilon_{1}}}
{\left(s_{f}- s_{0}+\varepsilon_1 +\varepsilon_{2}\right)},
\eea
which holds generically with the logarithmic accuracy for   
 $\varepsilon_1, \varepsilon_2\to 0$. Applying it repeatedly, we finally get
\be
\int_{s_0}^{s_f}\D_{\rm diff}^{(N)} s  \stackrel{\varepsilon\ll1}
\longrightarrow \frac{N}{\ln \frac 1\varepsilon}
\frac{1}{\left(s_{f}- s_{0} \right)}
\label{propae}
\ee
and
\be
\LA \left(\frac{(s_{K+1}-s_K)(s_{K}-s_{K-1})}{(s_{K+1}-s_{K-1})}\right)^a \RA
 \stackrel{\varepsilon\ll1}
\longrightarrow  \frac 1 {N\ln \frac1\varepsilon}
\frac{\Gamma^2(a)}{\Gamma(2a)} (s_f-s_0)^a \,.
\label{asympt}
\ee

In the derivation of these formulas it was saliently assumed that $N \ll \ln \frac 1\varepsilon$ in order to substitute
\be
\ln^N \frac {C}{\varepsilon}= \ln^N \frac {1}{\varepsilon}+{\cal O}
\left(\ln^{N-1} \frac {1}{\varepsilon}\right),
\ee
but they may have a wider domain of applicability. Anyway, comparing Eqs.~\rf{propa} and \rf{compli} with Eqs.~\rf{propae} and \rf{asympt}, we conclude that $1/\ln \frac 1\varepsilon$ for the logarithmic subordinator plays, roughly speaking, a role similar to $\delta$ for the gamma subordinator.

With this accuracy it can be shown that the $n$-step transition probability
\be
P\left(\Delta s,n\right)=\frac{n}{\ln ^n\frac 1\varepsilon}
\frac{\ln ^{n-1}\frac {C_n\Delta s}{\varepsilon}}{(\Delta s + n\varepsilon)},
\qquad P\left(\Delta s,1\right)\equiv P\left(\Delta s\right)
\label{n-step}
\ee
satisfies the Chapman--Kolmogorov chain condition
\be
P\left(s_{i+n}-s_{i-1},n+1\right) = \int_{s_{i-1}}^{s_{i+n}} \d s_i\,
P\left(s_{i+n}-s_i,n\right)P\left(s_i-s_{i-1}\right)
\label{CK}
\ee
and is associated with a stable subordinator process. This can be demonstrated calculating the characteristic function given by the Laplace transform
\be
\int_{0}^\infty \d x P(x,n) \e^{-q x} =
\left(\frac{\ln (\varepsilon q)}{\ln \varepsilon}   \right)^n
\label{LaTraLog}
\ee
with the logarithmic accuracy which requires $q\varepsilon \ll 1$.

\section{Numerical simulations\label{s:3}}

\subsection{The set up}

 In order to simulate the path integral over reparametrizations, we apply the Metropolis--Hastings algorithm with the weight function \rf{measure}. Note first that the distributions \rf{gamma-s} and \rf{log-s} cannot be normalized on the interval $\Delta s \in [0, +\infty)$, which makes the usual simulations of free random walks impossible. That is, one cannot simulate the subordinated process $s_{i}$ by generating random $\Delta s_{i}$ with the probabilities \rf{gamma-s} or \rf{log-s} and adding them to obtain $s_{i}$. Therefore, it is essential that $s_{i}$ satisfies the boundary conditions \rf{boundaryconds}, which provides an ``infrared cutoff'' for the distributions \rf{gamma-s} and \rf{log-s} at large $\Delta s$ and makes the problem well-posed.

 In the limits $\delta \rightarrow 0$ or $\varepsilon \rightarrow 0$, the distributions also become singular at small $\Delta s$. The measure \rf{measure} thus favors trajectories, for which all $s_{i}$ are very close to each other or equal. However, since the trajectories are subject to the boundary conditions \rf{boundaryconds}, some $\Delta s_{i}$ should be large. As we shall see, typical trajectories for the measure \rf{measure} consist of pieces with all $s_{i}$ almost equal and a finite number of discontinuous jumps with $\Delta s_{i} \sim s_{f} - s_{0}$. Since the typical size of such jumps is determined only by the ``infrared cutoff'' $(s_{f} - s_{0})$, they are somewhat similar to L\'evy's flights (see, e.g.\ Ref.~\cite{App04}), which are typical for random walks with infinite dispersion of the step probability distribution. This is in contrast to the Brownian trajectories which occur for nonsingular distributions -- if one simulates such an $N$-step subordinated process with the boundary conditions \rf{boundaryconds}, all steps $\Delta s_{i}$ will be of order of $(s_{f} - s_{0})/{N}$ and there will be no discontinuous jumps with $\Delta s_{i} \sim s_{f} - s_{0}$.
It is important to stress that our problem of a subordinator that hits a boundary is quite different from the usual free random walks, where each step is not restricted to be positive. Nevertheless, for nonsingular $P(\Delta s_i)$ in \rf{measure}, typical trajectories will be continuous, with each step $\Delta s_{i} \sim 1/N$, in spite of the fact that our measure does not involve terms like $\exp\left( -\! \int \dot{x}^2 \right)$.

\begin{figure}
\vspace*{3mm}
\includegraphics[width=9cm]{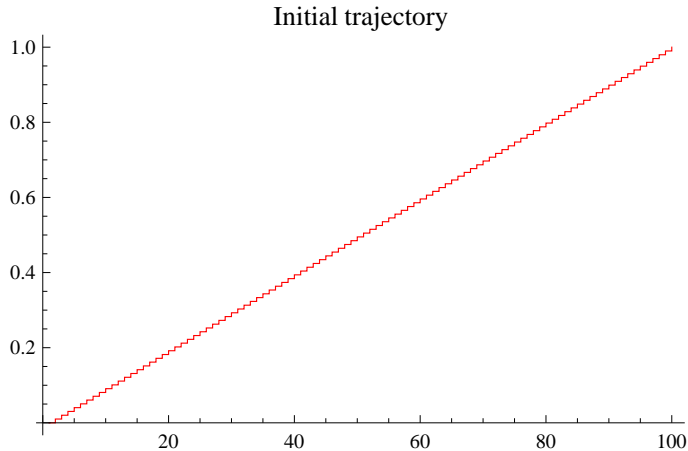}
\caption[]{Initial trajectory $s_{i}$ to apply the Metropolis--Hastings algorithm.}
\label{fi:initial}
\end{figure}

In order to apply the Metropolis--Hastings algorithm, we start from the initial trajectory of the type depicted in Fig.~\ref{fi:initial},
which is formed by small regular steps $=1/N$, and generate $s_i$ randomly distributed between $s_{i-1}$
and $s_{i+1}$. This new value is accepted if the inequality
\be
\frac{P\left( s_{i+1}-s_i \right)P\left( {s_i}-s_{i-1} \right)}
{P\left( s_{s+1}-s_i({\rm old}) \right)P\left( {s_i}({\rm old})-s_{i-1} \right)}
>r
\ee
is satisfied for random $0\leq r \leq1$. The averages like in \eq{avera} are then given by arithmetic means over an equilibrium ensemble of trajectories.

 It is also interesting to note that since the Metropolis--Hastings algorithm only involves the ratios of the measures \rf{measure}, it can be formally applied even to singular, non-normalizable distributions like \rf{gamma-s} and \rf{log-s} in the limits $\delta \rightarrow 0$ or $\varepsilon \rightarrow 0$. This fact should be contrasted, e.g., with the heatbath algorithm, which works only for normalizable distributions. However, in the case of a singular measure, the Metropolis--Hastings algorithm can spend arbitrarily large time near configurations with small $\Delta s$. Thus the singularities manifest themselves in an infinite slowdown of the algorithm. In practice, we have found that the autocorrelation time for the algorithm with the measure~\rf{log-s} grows as $\varepsilon^{-1}$ at small $\varepsilon \ll (s_{f} - s_{0})$.

\subsection{Hausdorff dimension\label{ss:3b}}

For $\delta \sim 1$ typical trajectories are similar to usual continuous trajectories of Gaussian random walks. Expanding the exact result~\rf{compliN} for large $N$, we obtain
\be
\LA \left(\frac{(s_{K+1}-s_K)(s_{K}-s_{K-1})}{(s_{K+1}-s_{K-1})}\right)^a \RA
\stackrel{N\gg1}\propto N^{-a}.
\label{compliN1}
\ee
An analogous result holds for a yet simpler average
\be
\LA \left(s_{K}-s_{K-1}\right)^a \RA =
\frac{\Gamma\left( N\delta \right)}{\Gamma\left( a+N\delta \right)}
\frac{\Gamma\left( a+\delta \right)}{\Gamma\left( \delta \right)
}\stackrel{N\gg1}\to N^{-a}\,.
\label{yetsimpler}
\ee
It is easy to understand this dependence to be typical for usual continuous trajectories, when $s_1-s_{i-1} \sim 1/N$, because
\be
\LA \left(\frac{(s_{K+1}-s_K)(s_{K}-s_{K-1})}{(s_{K+1}-s_{K-1})}\right)^a \RA
\stackrel{N\gg1}\to N^{-a}
\LA \left( \frac{s'(\tau)}{2}  \right)\RA_{\rm cont}\,.
\ee
Here $\tau=K/N$ is the value of the (normalized) proper time at the point, where the derivative $s'(\tau)$ is inserted. This scaling behavior of the gamma subordinator corresponds~\cite{Hor68} to the Hausdorff dimension equal to one.

On the contrary the result~\rf{compli} in the opposite limit of $N\to\infty$, when $N\delta\to 0$, is $N^{1-a}$ times smaller than the usual one. This holds also for the yet simpler average~\rf{yetsimpler}:
\be
\LA \left(s_{K}-s_{K-1}\right)^a \RA \stackrel{N\delta\ll1}=
\frac{\left(s_f-s_0 \right)^a}N\,,
\label{1N}
\ee
where the coefficient does not depend on $a$. This latter result can be understood as follows. In the numerical computations the average \rf{yetsimpler} is represented by
\be
\LA \left(s_{K}-s_{K-1}\right)^a \RA =
\LA \frac{1}{N} \sum_{i=1}^{N} \left(s_{i}-s_{i-1}\right)^a  \RA_{\rm config}.
\label{cosum}
\ee
Let us assume that a given configuration has only one flight with the magnitude $(s_f-s_0)$ that happens at an arbitrary point $i$. Then \eq{1N} is precisely reproduced.

Such a configuration does not contribute, however, to the average~\rf{compli} because the summand in
\be
\LA \left(\frac{(s_{K+1}-s_K)(s_{K}-s_{K-1})}{(s_{K+1}-s_{K-1})}\right)^a \RA
=\LA \frac{1}{N-1} \sum_{i=1}^{N-1}
\left(\frac{(s_{i+1}-s_{i})(s_{i}-s_{i-1})}{(s_{i+1}-s_{i-1})}\right)^a
\RA_{\rm config}
\label{cosumi}
\ee
would vanish. A nonvanishing contribution comes from those configurations for which a flight happens in two consecutive steps with the magnitudes $x$ and $1-x$ ($0<x<1$). An integration over $x$ reproduces the gamma-functions in \eq{compli}, while the factor of $\delta$ is apparently due to the fact that the probability for a flight to happen in two steps is a factor of $\delta$ smaller than to happen in one step. This can be demonstrated by calculating
\begin{subequations}
\bea
\LA (s_K-s_{K-1}) \RA &=&\frac{1}N (s_f-s_0), \\
\LA (s_{K+1}-s_K)(s_{K}-s_{K-1}) \RA &=&\frac{\delta}{N(1+N\delta)} (s_f-s_0)\,.
\eea
\end{subequations}
The second formula shows the difference between the scaling regimes for $N\to \infty$, $\delta \sim 1$ and $N\to \infty$, $N \delta \to 0$. These two scaling regimes correspond to the Hausdorff dimensions one and zero, respectively. 

A standard way~\cite{Hor68} to calculate the Hausdorff dimension for a stochastic process is based on the L\'evy--Khintchine representation of the characteristic function
\be
\int _0^\infty \d x P(x,\tau) \e^{-q x} = \e^{-\tau g(q)}
\label{34}
\ee
and is given by the formula
\be
d_H= \lim_{q\to\infty} \frac{\ln g(q)}{\ln q}\,.
\label{DH}
\ee
For the standard gamma-subordinator we have
\be
P(x,\tau) = \frac{b^\tau}{\Gamma(\tau)} x^{\tau-1} \e^{-b x}
\label{36}
\ee
and
\be
\e^{-\tau g(q)}= \left(1+\frac q b \right)^{-\tau} \,,
\ee
so that
\be
d_H=\lim_{q\to\infty}\frac{\ln \ln \left(1+\frac q b \right)}{\ln q}.
\label{DHs}
\ee
This gives $d_H=0$ for $b\sim 1$.

For our process with discrete time $\tau=\delta N$, we supersede \eq{34} by 
\be
\LA \e^{-q(s_{K+n}-s_K)}\RA = \e^{-n g (q)/N}
\label{defg}
\ee
and adopt the definition~\rf{DH} of the Hausdorff dimensions,
whose discrete analog explicitly reads as
\be
d_H=\lim_{q\to\infty}\frac{\ln \left[-\frac Nn
\ln \LA \e^{-q(s_{i+n}-s_i)} \RA \right]}{\ln q}.
\label{defDH2}
\ee
It can be shown following Ref.~\cite{Hor68} that in the limit $N\to\infty$ this definition is equivalent to the geometric one, obtained by covering the one-dimensional set of points $s_i$'s by intervals.

\begin{figure}
\vspace*{3mm}
\includegraphics[width=7.8cm]{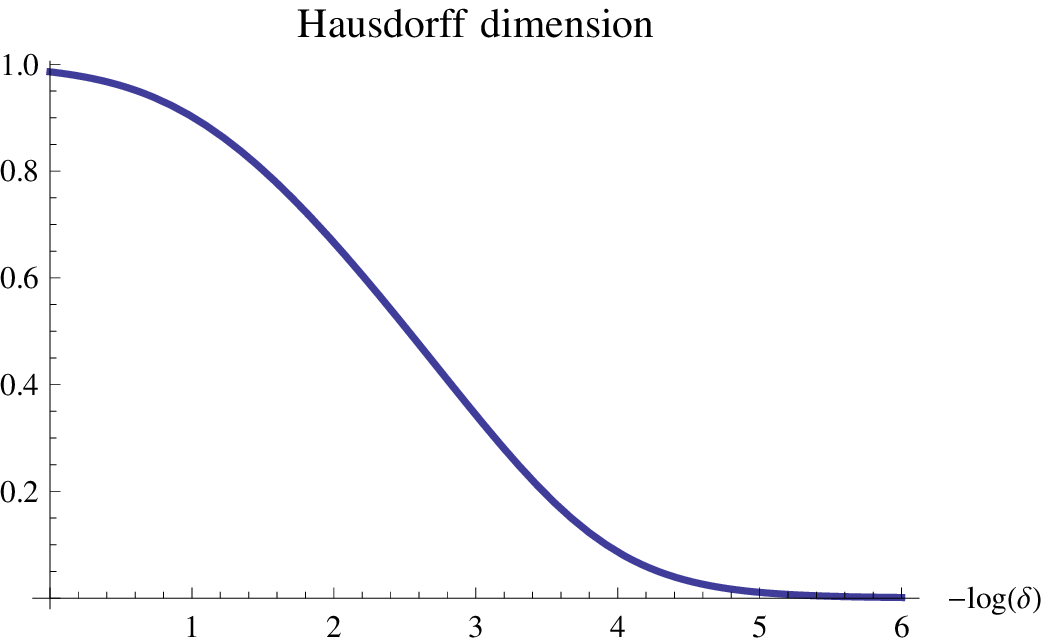} \hspace*{4mm}
\includegraphics[width=7.8cm]{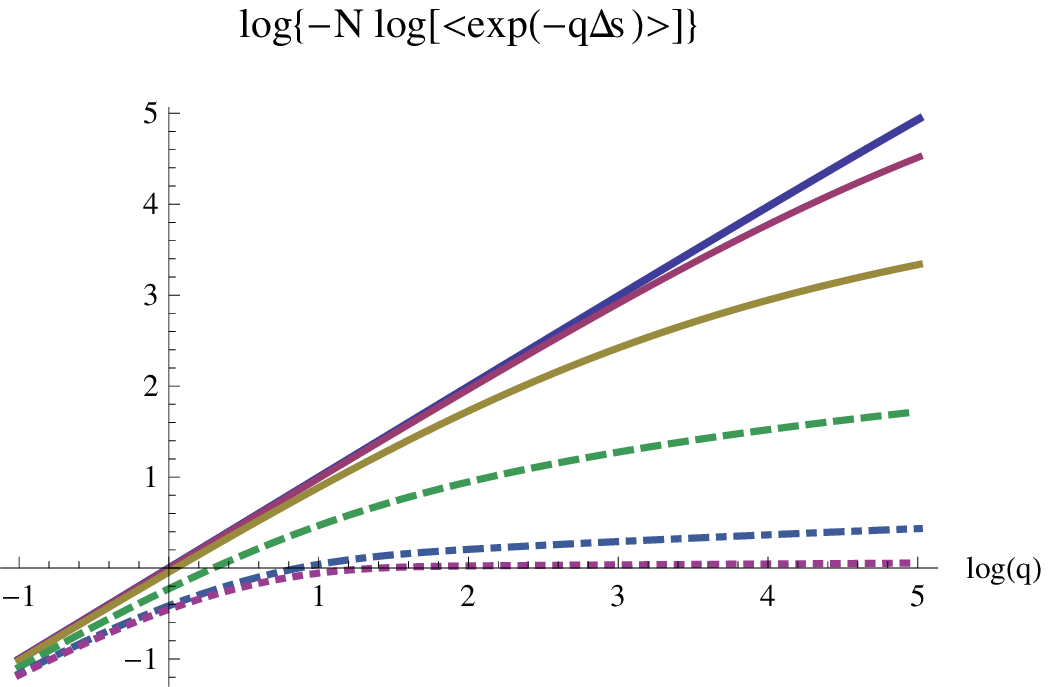}
 \caption[]{Hausdorff dimension \rf{defDH2}, \rf{1F1} versus  $\ln(1/\delta)$ (left) extracted from the behavior of $g(q)$ for different $\delta$ (right) at $N=1000$. The lines in the right figure corresponds to $\delta=1, 10^{-1}, 10^{-2}, 10^{-3}, 10^{-4}, 10^{-5}$ from the top to the bottom. Their slope equals the Hausdorff dimension.}
\label{fi:HDgs}
\end{figure}

Generalizing \eq{yetsimpler}, we obtain for \rf{defg} the confluent hypergeometric function
\be
\e^{-n g (q)/N}= \frac{\Gamma(\delta N)}{\Gamma(\delta n)}
\sum_{k=0}^{\infty} \frac{(-q)^{k}}{k!}
\frac{\Gamma(k+\delta n)}{\Gamma(k+\delta N)}
={}_1F_1(\delta n, \delta N;-q)
\label{1F1}
\ee
that determines at large $N$
\be
\e^{-n g (q)/N} =\left\{
\begin{array}{ll}
\left(1+\frac q{\delta N} \right)^{-\delta n}
\stackrel{N\delta\gg 1}\to\e^{-n q/N} \qquad &\hbox{for}~~\delta\sim 1 \\
1+\frac{n}{N}\left(\e^{- q} -1\right)\qquad& \hbox{for}~~N\delta\ll1
\end{array}
\right. \,.
\label{(30)}
\ee
After the substitution into the definition \rf{defDH2}, \eq{(30)} reproduces $d_H=1$ and $d_H=0$, correspondingly. This value of the Hausdorff dimension versus $\ln(1/\delta)$ is plotted in Fig.~\ref{fi:HDgs} at $N=1000$.
It is clear from the figure that the desired $N\delta\to0$ continuum limit is reached for $\delta \leq 10^{-5}$. If the value of $N$ is smaller, this limit sets in for yet larger values of $\delta$. Analogously, \eq{LaTraLog} for the characteristic function of the logarithmic subordinator in the logarithmic approximation results in $d_H=0$.

A convenient quantity, describing the trajectories, is the spectral density
\be
\rho(x,n)\equiv \LA \delta^{(1)} \left( s_{i+n}-s_i-x \right) \RA
\ee
which describes how many $\Delta s_i$'s takes the value of $x$. Using \eq{(30)}, we find
\be
\LA \delta^{(1)} \left( s_{i+n}-s_i-x \right) \RA  =\left\{
\begin{array}{ll}
\delta N \e^{-x \delta N} (x\delta N)^{\delta n -1}/\Gamma (\delta n)
\qquad &\hbox{for}~~N\delta\to \infty \\
\left(1-\frac n N\right)
\delta^{(1)}\left(x\right)+\frac n N\, \delta^{(1)} \left(1-x\right)
\qquad& \hbox{for}~~N\delta\to 0
\end{array}
\right. \,.
\label{(31)}
\ee
These are two limiting cases of the exact formula
\be
\rho(x,n)=\frac{\Gamma(N\delta)}{\Gamma(n\delta)\Gamma\left((N-n)\delta\right)}
x^{n\delta-1} (1-x)^{(N-n)\delta-1}\,.
\label{rhoexact}
\ee
If additionally $n\delta\to\infty$, the density~\rf{rhoexact} has a very sharp peak at $x=n/N$:
\be
\rho(x,n)\stackrel{n\delta\to\infty} \longrightarrow \delta^{(1)}\left(
x-\frac nN \right),
\ee
which is associated with the trajectory of the type in Fig.~\ref{fi:initial}.

An exact analog of \eq{rhoexact} exists also for an arbitrary distribution:
\be
\rho(x,n)=(s_f-s_0)
\frac{P\left(x(s_f-s_0),n \right)P\left((1-x)(s_f-s_0),N-n \right)}
{P\left((s_f-s_0),N \right)},
\ee
which is normalized to 1 as a consequence of the Chapman--Kolmogorov equation~\rf{CK}.

\subsection{Monte-Carlo results}

\begin{figure}
\vspace*{3mm}
\includegraphics[width=8cm]{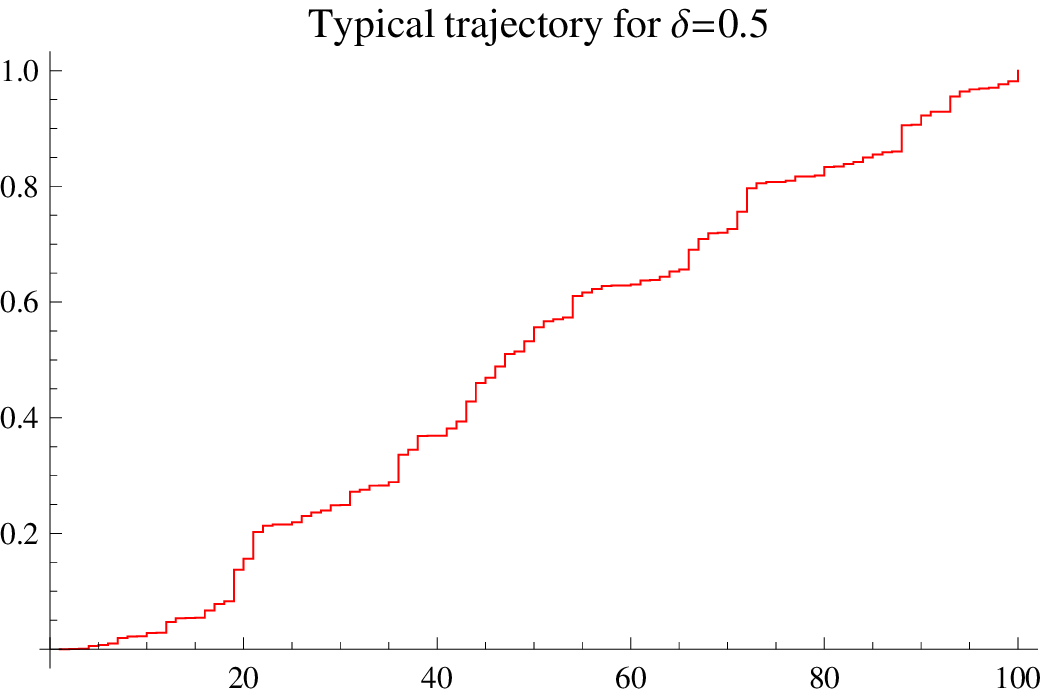}
\includegraphics[width=8cm]{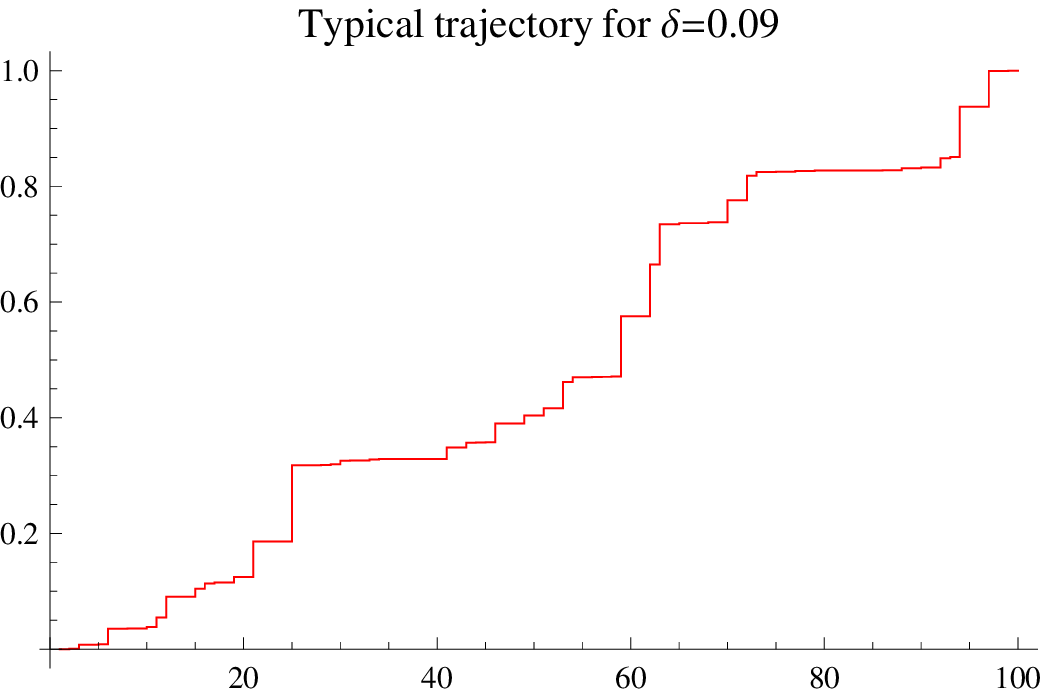}
\caption[]{Typical trajectories $s_{i}$ for PDF~\rf{gamma-s} with $\delta=0.5$ (left) and $\delta=0.09$ (right).}
\label{fi:typical05gamma}
\end{figure}

Some typical trajectories for the PDF~\rf{gamma-s} are depicted in Fig.~\ref{fi:typical05gamma} for $\delta=0.5$ and $\delta=0.09$.
The trajectory for $\delta=0.5$ looks rather normal, while the L\'evy flights appear for those at $\delta=0.09$.
The numerically calculated average approaches its exact value~\rf{compliN} after averaging over trajectories. For smaller $\delta$ the numerical algorithm does not apparently work well because PDF~\rf{gamma-s} is singular at $\Delta s_i=0$.

 Due to this singularity, it is very difficult to go to the required limit $N\delta \ll1$, so we switch to the logarithmic subordinator for which numerics works much better so we can go to the continuum limit for $N \varepsilon\ll1$, but analytical formulas are available only for very tiny $\varepsilon$.

\begin{figure}
\vspace*{3mm}
\includegraphics[width=8cm]{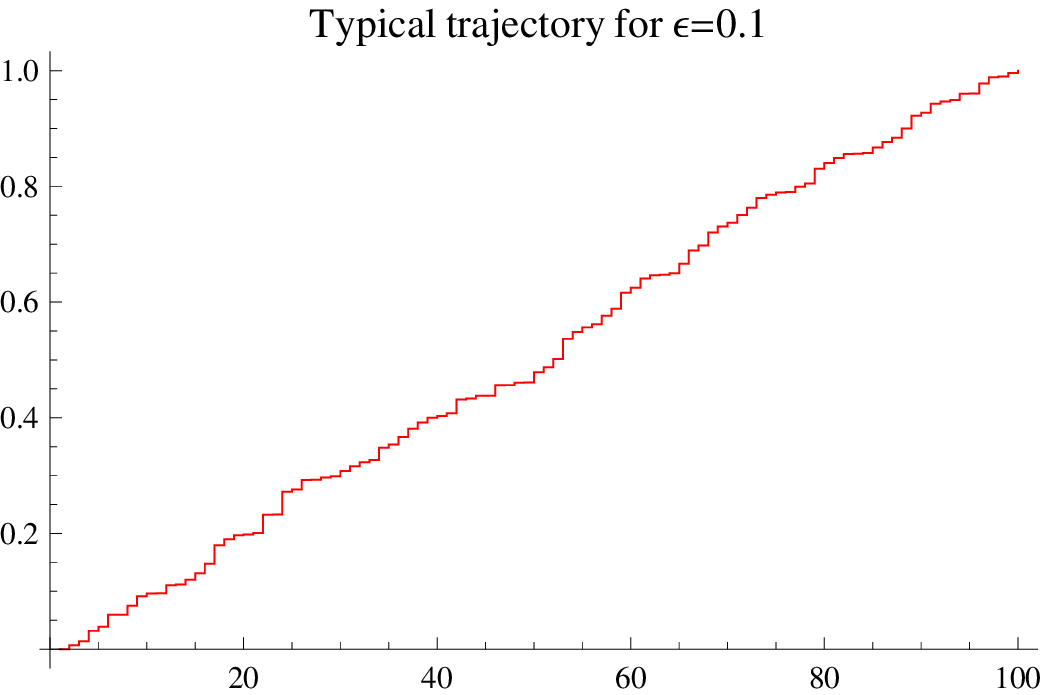}
\includegraphics[width=8cm]{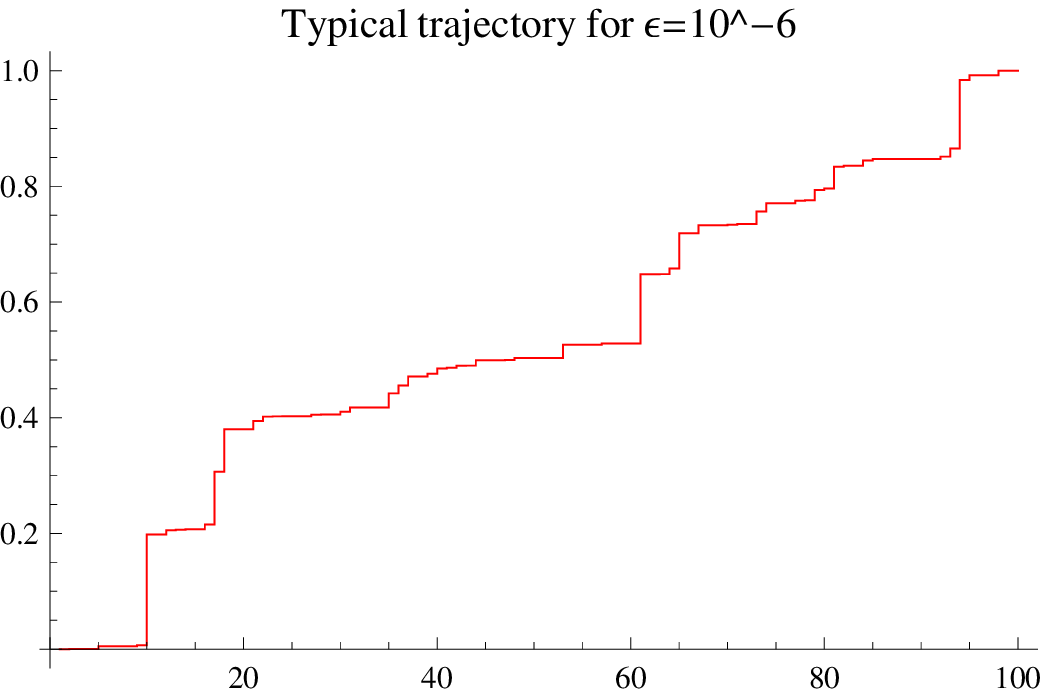}
\caption[]{Typical trajectories $s_{i}$ for PDF~\rf{log-s} for $\varepsilon=0.1$ (left) and $\varepsilon=10^{-6}$ (right).}
\label{fi:typicallog}
\end{figure}

The data are presented in Fig.~\ref{fi:typicallog} for $N=100$ and $\varepsilon=0.1$, $ 10^{-6}$.
The values of the averages~\rf{cosum} or \rf{cosumi} decrease with decreasing $\varepsilon$ as is expected, but are much larger than ones given by \eq{asympt}, so this asymptote apparently does not yet set in.

\begin{figure}
\vspace*{3mm}
\includegraphics[width=5.5cm,angle=-90]{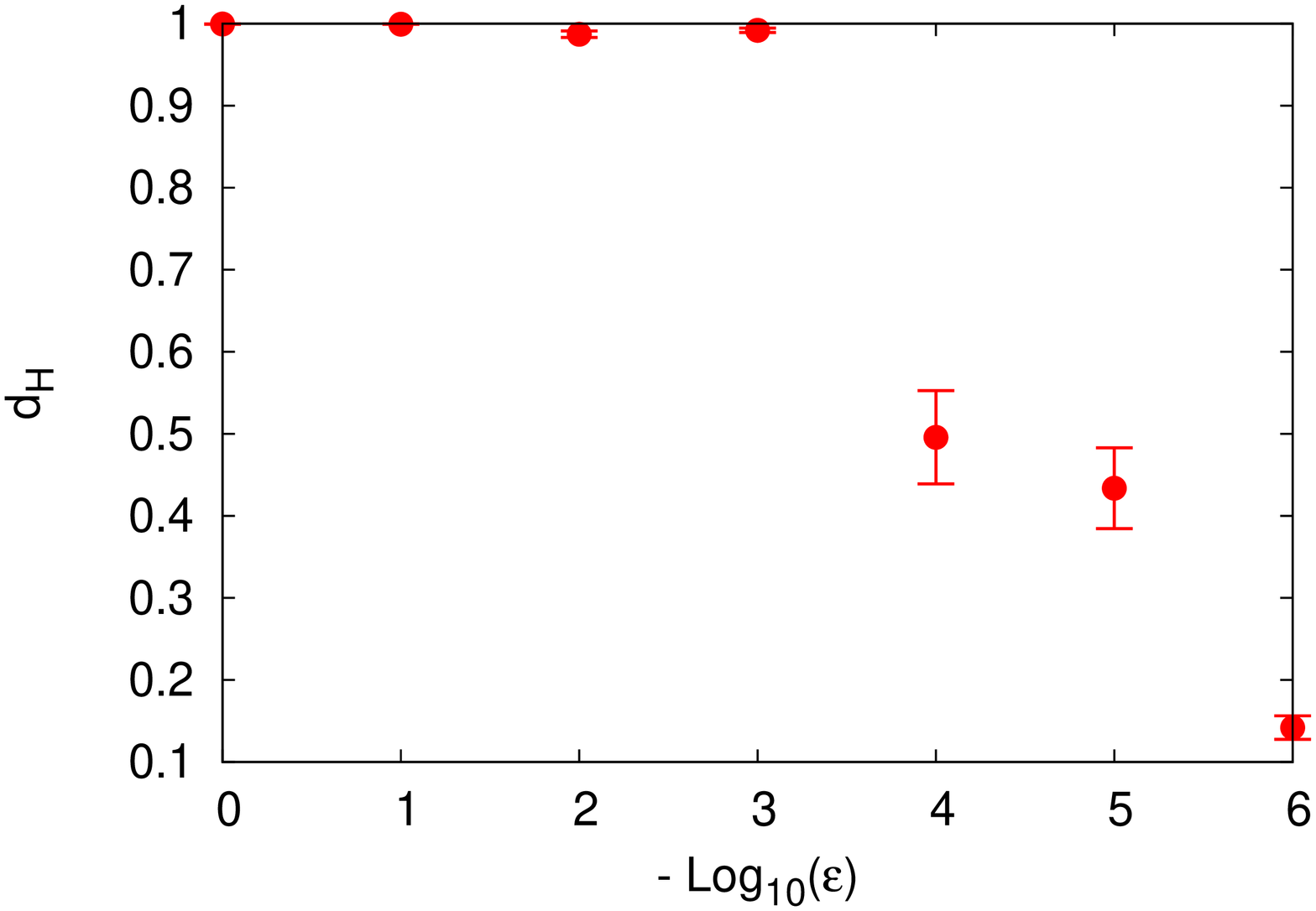}
\hspace*{5mm}
\includegraphics[width=5.5cm,angle=-90]{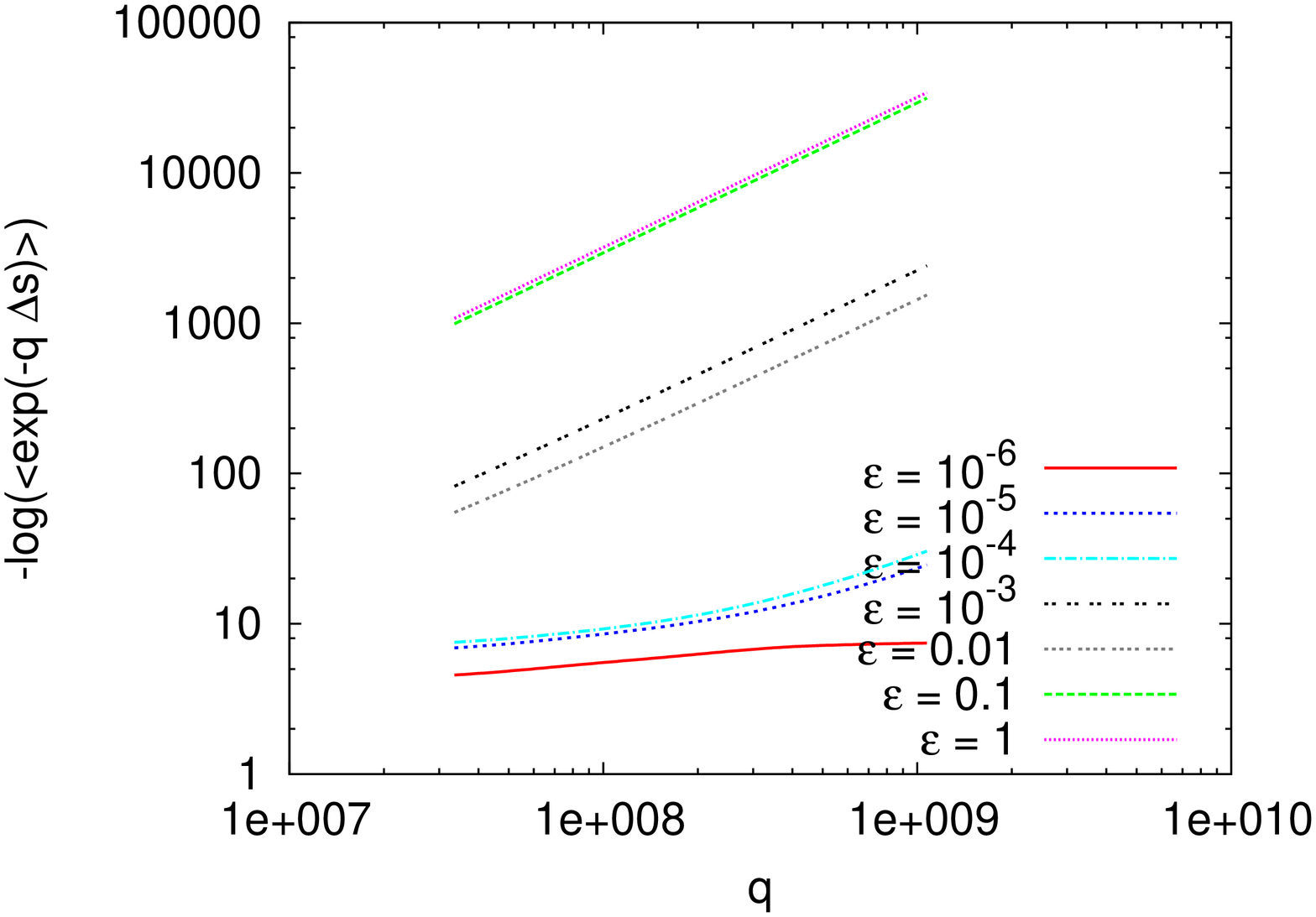}
 \caption[]{Hausdorff dimension of the logarithmic subordinator
versus  $\varepsilon$ (left) extracted from the behavior of $-\ln \LA\exp{-q \Delta s}\RA$ for different $\varepsilon$ (right) at $N = 50$.}
\label{fi:HDlog-s}
\end{figure}

The Hausdorff dimension of the logarithmic subordinator decreases, as expected, with decreasing $\varepsilon$ as is illustrated by Fig.~\ref{fi:HDlog-s}.


\section{Simulating effective string\label{s:4}}

We can numerically simulate the path integral over reparametrization in the effective string ansatz \cite{Pol97}
\be
\Psi(C) = \int \D_{\rm diff}s \e^{-K A[s]}
\label{Psi}
\ee
with the Douglas integral
\begin{equation}
 A[s]=\frac{1}{4\pi} \int_{-\infty}^{+\infty} \d t_1 \d t_2 \,
\frac{\left[ x(s(t_1)) - x(s(t_2))\right]^2}{(t_1-t_2)^2}
\label{altDouglas}
\end{equation}
for a particular $C$, say, for a circle or an ellipse. Then the exponential of  $-K A[s]$ is to be included in the measure. After an equilibrium set of configurations with this weight is generated, the averages of the type
\begin{equation}
\LA A[s] \RA \equiv \frac{\int \D_{\rm diff}s \e^{-K A[s]}A[s] }
{\int \D_{\rm diff}s \e^{-K A[s]}}
\label{defAa}
\end{equation}
can be calculated as arithmetic means.

From the definition \rf{Psi} we have
\be
\LA A[s] \RA = - \frac{\partial}{\partial K} \ln \Psi(C)\,.
\ee
For a circle of the radius $R$, $\Psi(C)$ has an expansion for large
\be
K R^2 \equiv \beta
\ee
of the form
\be
\Psi({\rm circle}) 
=\beta^{-\alpha} \e^{-\pi \beta}
\left(\,{\rm const.}+{\cal O}(1/\beta)\right),
\ee
which yields
\be
\LA \frac{A[s]}{R^2} \RA =- \frac{\partial}{\partial \beta}\ln \Psi(C)
=\pi  +\frac \alpha \beta + {\cal O}(1/\beta^2).
\ee
The approach to the asymptotic value $\pi$ is from above for $\alpha>0$.

For the sake of discretization it should be better to use a unit circle parametrization, changing the variable from
$s$ ($-\infty<s<+\infty$) to
\be
\sigma = -2 {\rm arccot}~ s \qquad (0\leq \sigma \leq 2\pi)\,.
\ee
Discretizing the variable $\phi$ as
\be
\phi_i=\frac{2\pi i}N
\ee
and replacing $\sigma(\phi_i)\equiv \sigma_i$ by
\be
s_i=-\cot \frac{\sigma_i}2\,,
\label{ssig}
\ee
we obtain for a circle
\be
A\left[\{s_i\}\right]=\pi R^2\lim_{N\to\infty} \frac1{(N^2-N)} \sum_{i\neq j}
\frac{1}{{\sin^2[\pi(i-j)/N]}} \frac{(s_i-s_j)^2}{(1+s_i^2)(1+s_j^2)}
\ee
which is a (discrete version of the) functional of $s_i$'s over which we integrate with the same measure as before. The minimum of $A\left[\{s_i\}\right]$ is reached at
\be
s_i=-\cot\frac{\pi i}N\,,
\ee
when $A_{\rm min}[s_*]=\pi R^2$.

Above $s_i$ changes from $-\infty$  to $+\infty$. For the purposes of numerical simulations it is better to use the variable $\sigma \in [0,1]$, which is related to $s$ by \eq{ssig}. We rewrite \eq{altDouglas} for a circle as
\be
A[\sigma]= \frac{R^2}{4\pi} \int_0^{2\pi}\d \phi \int_0^{2\pi}\d \phi'\,
\frac{\sin^2\{[\sigma(\phi)-\sigma(\phi')]/2\}}
{\sin^2[(\phi-\phi')/2]} \,.
\label{Douglas1}
\ee
The discretized measure and the Douglas functional then read
\be
w[\{\sigma_i\}]=\frac{1}{\sin[\pi(\sigma_{i}-\sigma_{i-1})]+\varepsilon}
\exp\left(-\beta A[\{ \sigma_i\}]\right),
\label{measureD}
\ee
\be
A[\{ \sigma_i\}]=
\pi R^2\lim_{N\to\infty} \frac1{(N^2-N)} \sum_{i\neq j}
\frac{\sin^2[\pi(\sigma_i-\sigma_j)]}{\sin^2[\pi(i-j)/N]}\,.
\label{DouglasFunc}
\ee
The minimum of $A[\sigma]$ is then reached at
\be
\sigma_i=\frac{ i}N\,,
\ee
when $A[\sigma_*]=\pi R^2$.

\subsection{Monte-Carlo simulations of subordinated process with Douglas functional}

\begin{figure}
  \includegraphics[width=7cm, angle=-90]{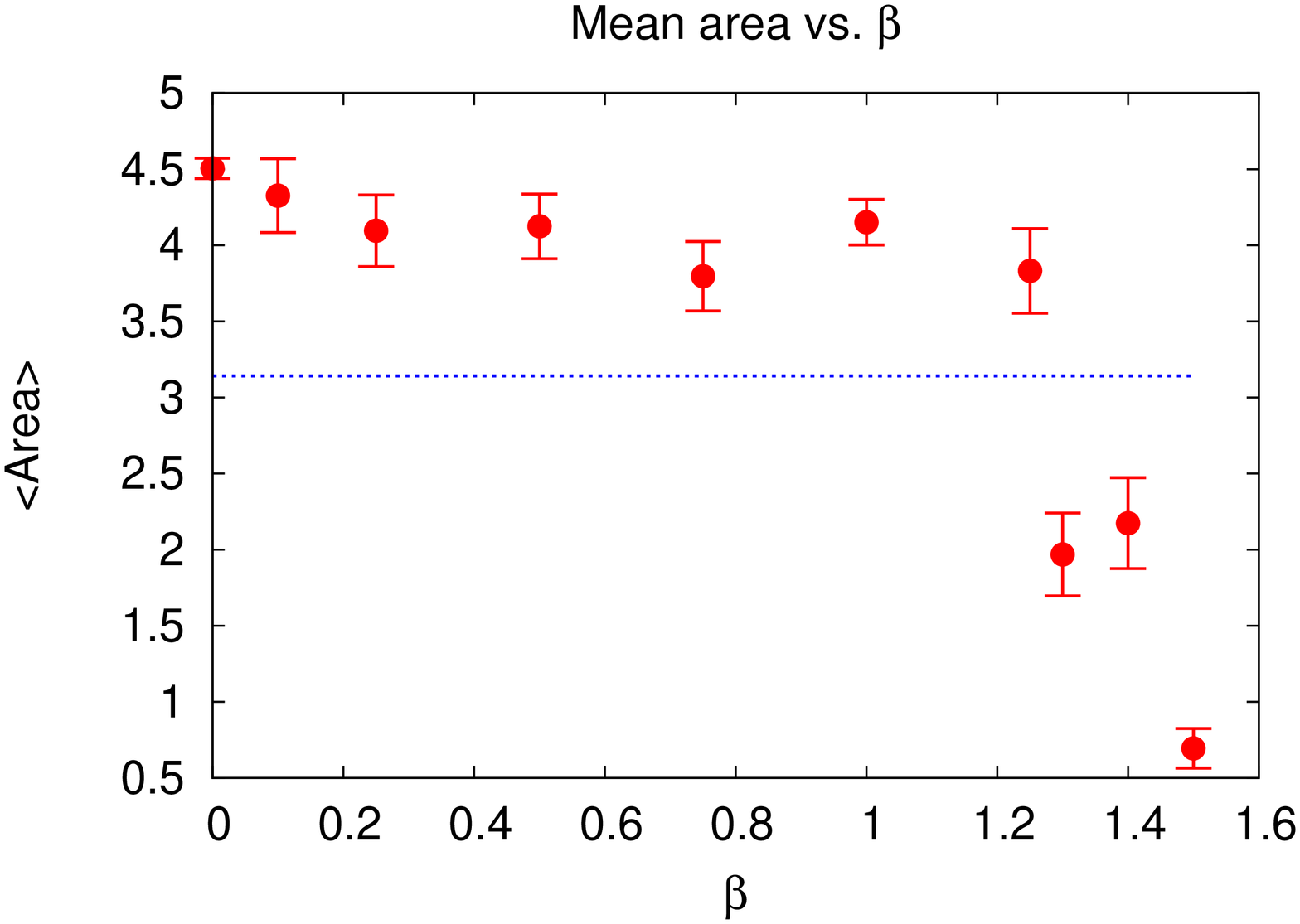}\\
   \caption{Mean value of the Douglas functional \rf{DouglasFunc} for a unit circle versus $\beta$ at $\varepsilon = 10^{-6}$ and $N = 50$. The solid line represents the minimal area $A_{\rm min}[s] = \pi$.}
\label{fi:area_vs_beta}
\end{figure}

\begin{figure}
\vspace*{3mm}
\includegraphics[width=7cm]{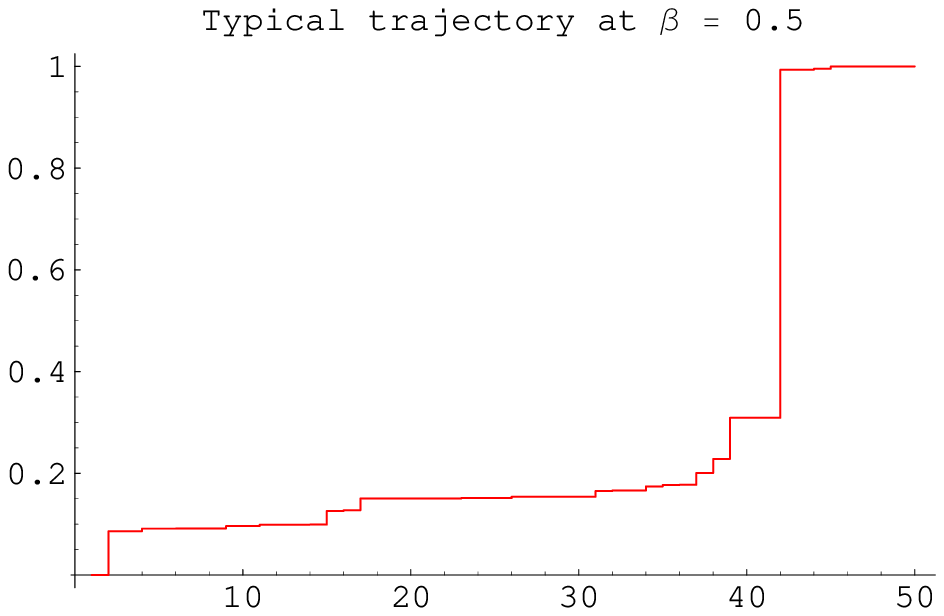}
\includegraphics[width=7cm]{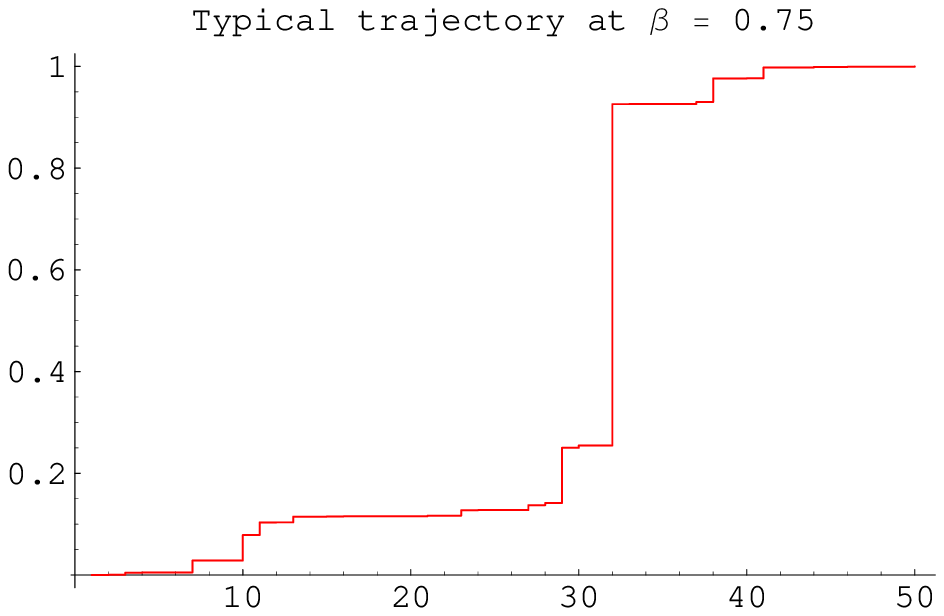}\\[.5cm]
\includegraphics[width=7cm]{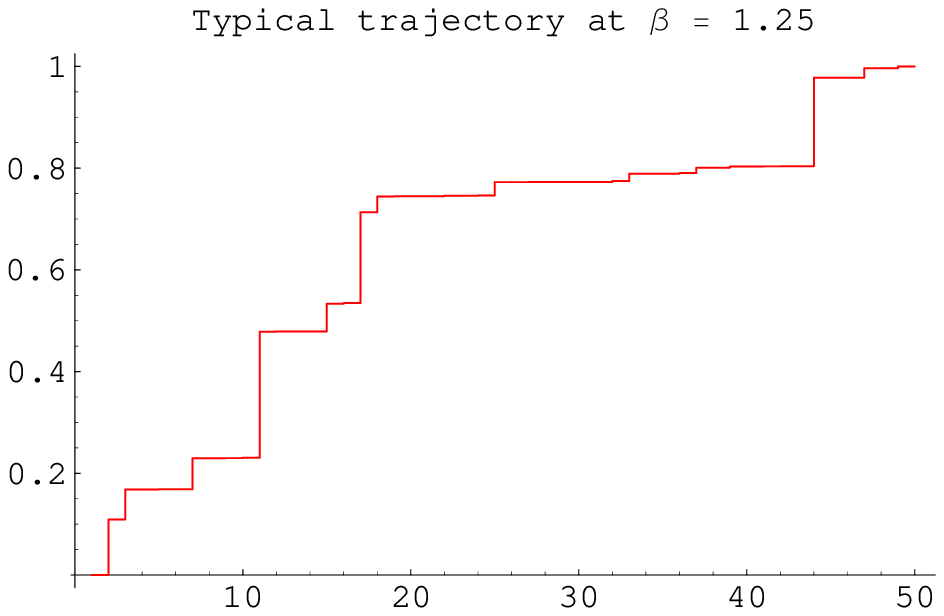}
\includegraphics[width=7cm]{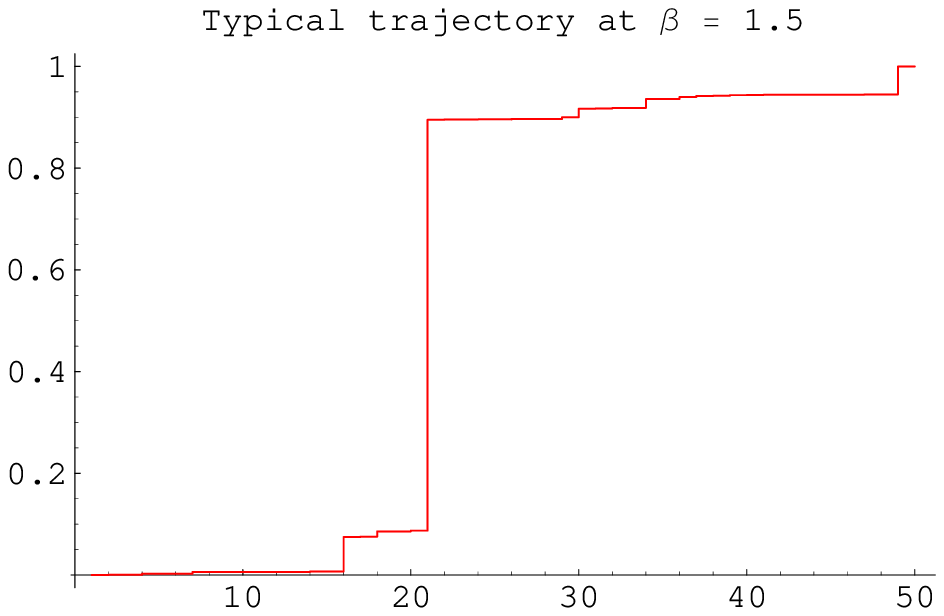}\\
\caption[]{Typical trajectories $s_{i}$ for the measure \rf{measureD} for $\varepsilon = 10^{-6}$ at various $\beta$.}
\label{fi:typicallogbeta}
\end{figure}

\begin{figure}
\vspace*{3mm}
\includegraphics[width=5.5cm,angle=-90]{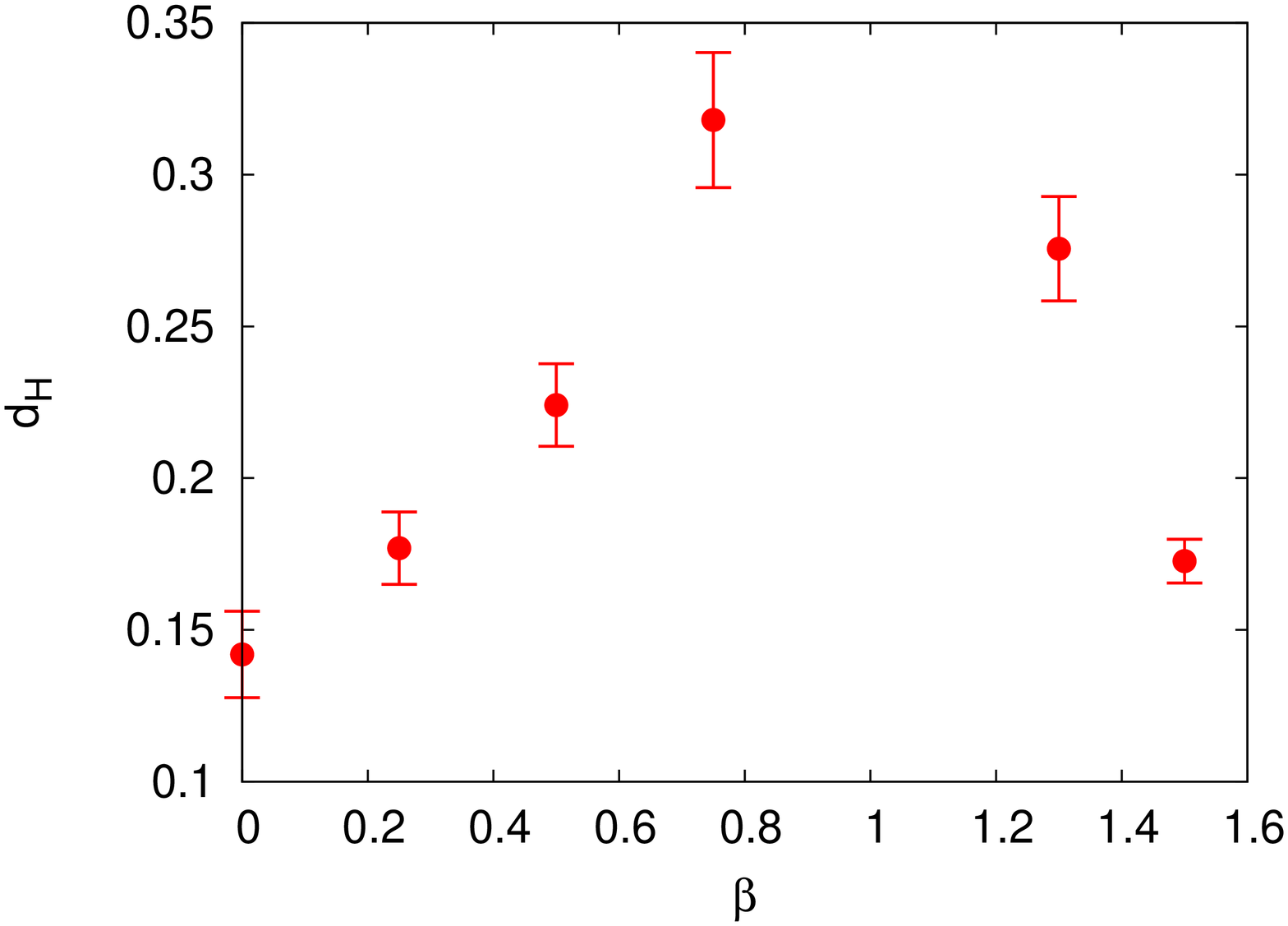}
\hspace*{5mm}
\includegraphics[width=5.5cm,angle=-90]{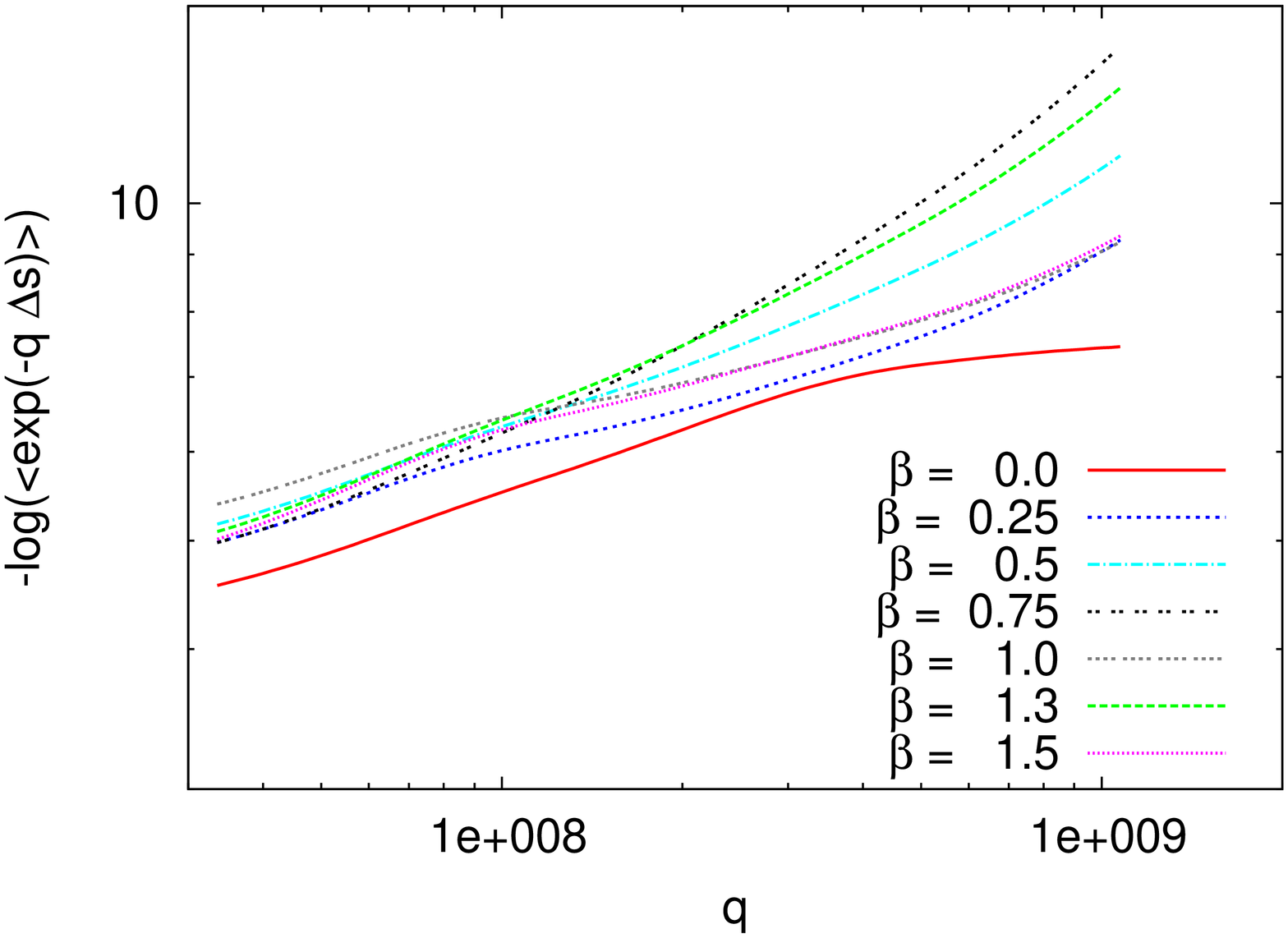}
 \caption[]{Hausdorff dimension of the subordinator with the measure \rf{measureD} versus $\beta$ (left) extracted from the behavior of $-\ln \LA\exp{-q \Delta s}\RA$ at $\varepsilon = 10^{-6}$ (right).}
\label{fi:HDlog-s-beta}
\end{figure}

 We have performed Monte-Carlo simulations of subordinated trajectories with the measure \rf{measureD}, using the Metropolis--Hastings algorithm. The average area\footnote{That is, the average value of the discretized Douglas functional \rf{DouglasFunc} over an equilibrium ensemble of subordinators $s_{i}$.} as a function of $\beta$ at $\varepsilon = 10^{-6}$ and $N = 50$ is plotted in Fig.~\ref{fi:area_vs_beta}. Typical trajectories are plotted in Fig.~\ref{fi:typicallogbeta}, and their Hausdorff dimensions are shown in Fig.~\ref{fi:HDlog-s-beta}. One can see that for small $\beta$ the mean area decreases with $\beta$, as expected. Typical trajectories become somewhat more smooth, and their Hausdorff dimension increases up to $d_{H} \approx 0.3$. However, at $\beta \approx 1.2$ the average area abruptly decreases below the minimal value allowed for continuous trajectories. Correspondingly, the jumps in the typical trajectories appear again, and their Hausdorff dimension decreases.

The reason for such an anomalous behavior is that Douglas' theorem states that the minimal area, spanned by a given loop, is given by the absolute minimum of the Douglas functional in the space of all \emph{continuous reparametrizations}, which are smooth. For a discretized functional \rf{DouglasFunc} this theorem is apparently not applicable. In particular, one can consider a degenerate trajectory with $s_{i} = 0$ if $i \le k$ and $s_{i} = 1$ if $i > k$ for some $k$. For such trajectories the discretized Douglas functional \rf{DouglasFunc} is identically equal to zero! The discretized classical trajectory with $s_{i} = i/N$ is still a minimum, as one can easily check by solving the extremum condition, but only a local one. The global minimum $A[s] = 0$ corresponds to a trajectory with a single jump from $0$ to $1$.

\subsection{Path integral over reparametrizations with (improved) Douglas functional}

 As we have seen in the previous Subsection, the naive discretization of the Douglas functional conflicts with Douglas' theorem. In particular, the global minimum of thus discretized functional equals zero rather than the minimal area spanned by the loop. The reason for such an anomalous behavior can be clearly identified: Douglas' theorem is valid only for continuous reparametrizations with nonsingular first derivative, while discontinuities of the type of unit jumps lead to wrong values of the area. To avoid this problem, we can approximate the reparametrization by a continuous polygonal line rather than by discrete steps:
\begin{eqnarray}
 \sigma\left(\tau\right) = \sigma_{i} + (\sigma_{i+1} - \sigma_{i}) \, 
\left(N\tau - i\right), 
 \quad \frac{i}{N} < \tau < \frac{i+1}{N}\,.
\label{ReparametrDiscrNew}
\end{eqnarray}
The Douglas functional can then be represented as
\begin{eqnarray}
 A[\{ \sigma_i\}]=
\pi R^2 \lim_{N\to\infty} \frac1{(N^2-N)} \sum_{i\neq j}
\int \limits_{\tau_i}^{\tau_{i+1}} d\tau_1 
\int \limits_{\tau_j}^{\tau_{j+1}} d\tau_2
\frac{\sin^2[\pi(\sigma\left(\tau_1\right)-\sigma\left(\tau_2\right))]}{\sin^2[\pi\left(\tau_1 - \tau_2\right)]}\,,
\label{DouglasDiscrNew}
\end{eqnarray}
where $\tau_{i} = i/N$ and $\sigma\left(\tau_1\right)$, $\sigma\left(\tau_2\right)$ are given by \rf{ReparametrDiscrNew}. The resulting integrals over the range $[\tau_{i}, \tau_{i+1}]$ can then be taken numerically by dividing this interval into some number $N'$ of intermediate points and applying a standard numerical integration method. The simulations would be also very efficiently accelerated if these integrals could be taken analytically. Unfortunately, this is not the case for the circle. Below we present the results obtained with thus ``improved'' discretization of the Douglas functional, which do not exhibit the anomalous behavior discussed in the previous Subsection.

\begin{figure}
  \includegraphics[width=7cm, angle=-90]{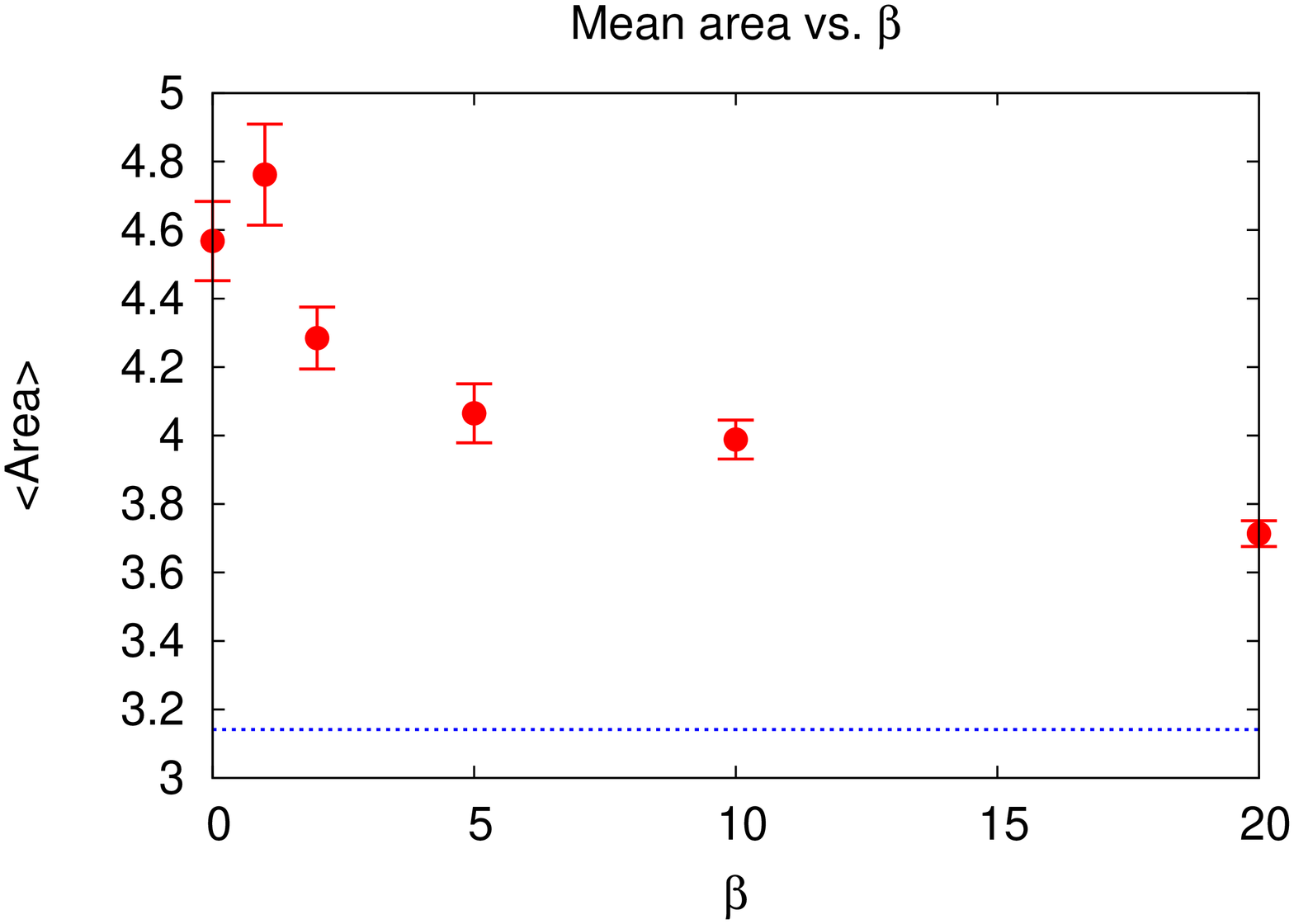}\\
   \caption{Mean value of the Douglas functional \rf{DouglasDiscrNew} for a unit circle versus $\beta$ at $\varepsilon = 10^{-6}$ and $N = 50$. The solid line represents the minimal area $A_{\rm min}[s] = \pi$.}
\label{fi:area_vs_beta_new}
\end{figure}

\begin{figure}
\vspace*{3mm}
\includegraphics[width=7cm]{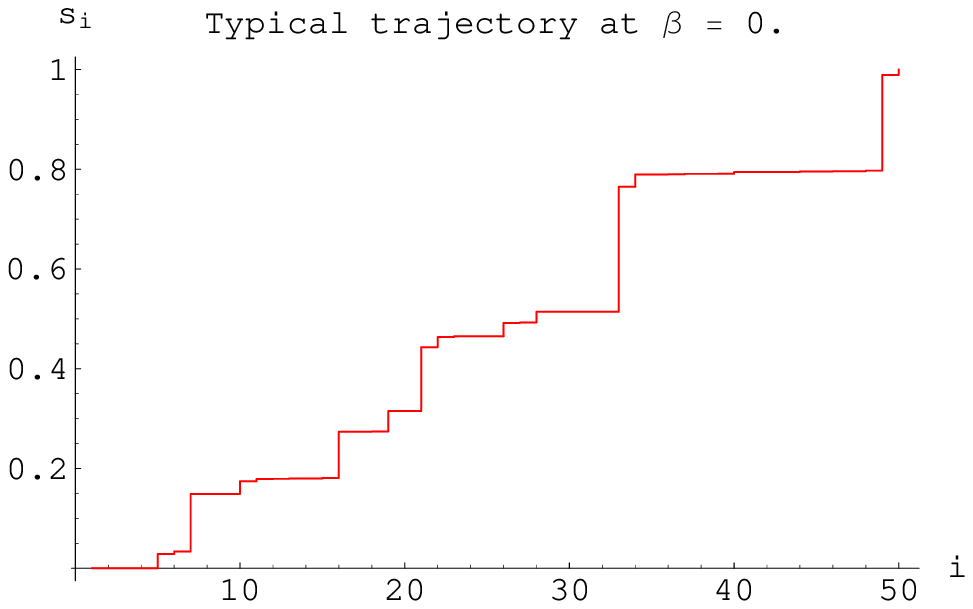}
\includegraphics[width=7cm]{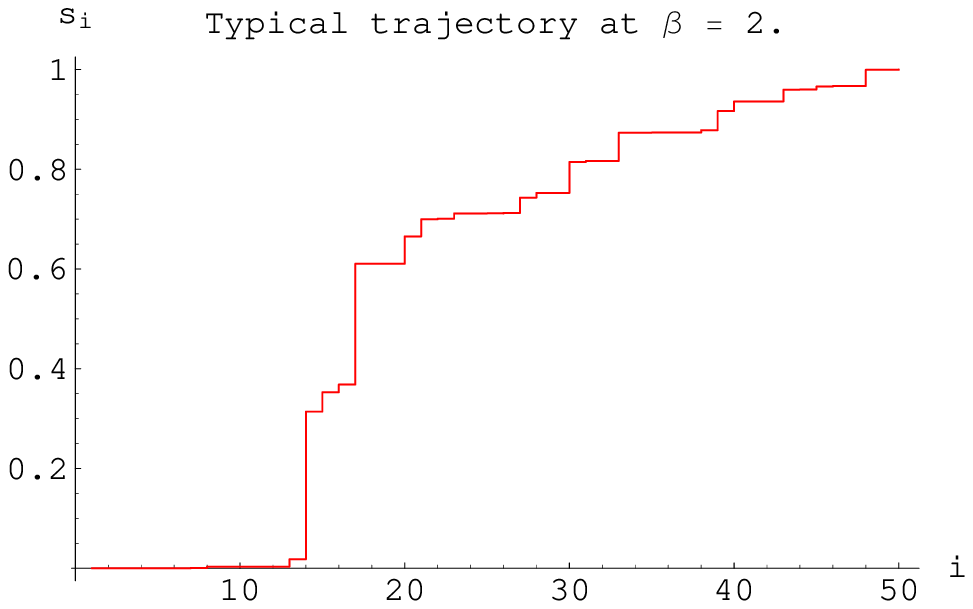}\\[.5cm]
\includegraphics[width=7cm]{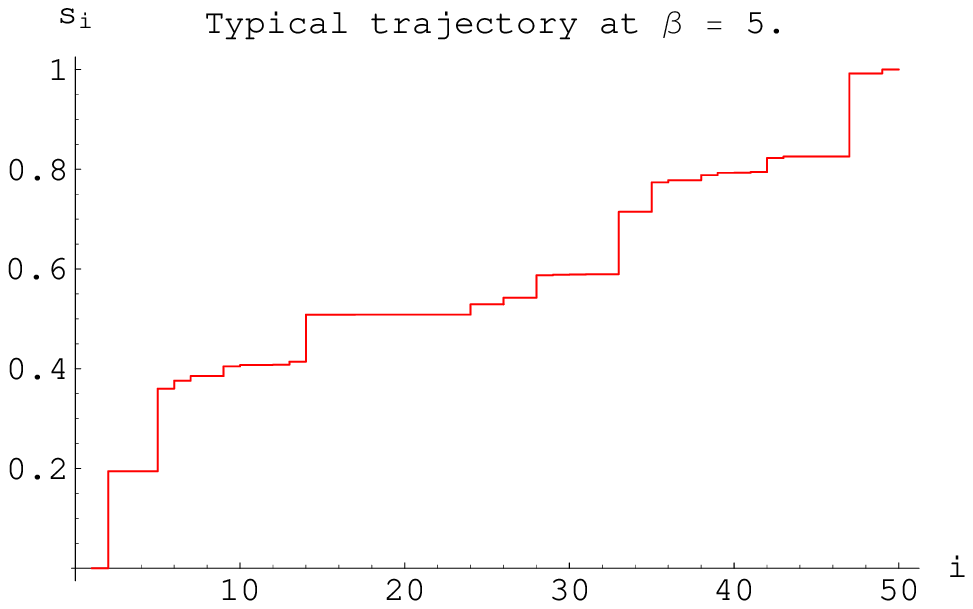}
\includegraphics[width=7cm]{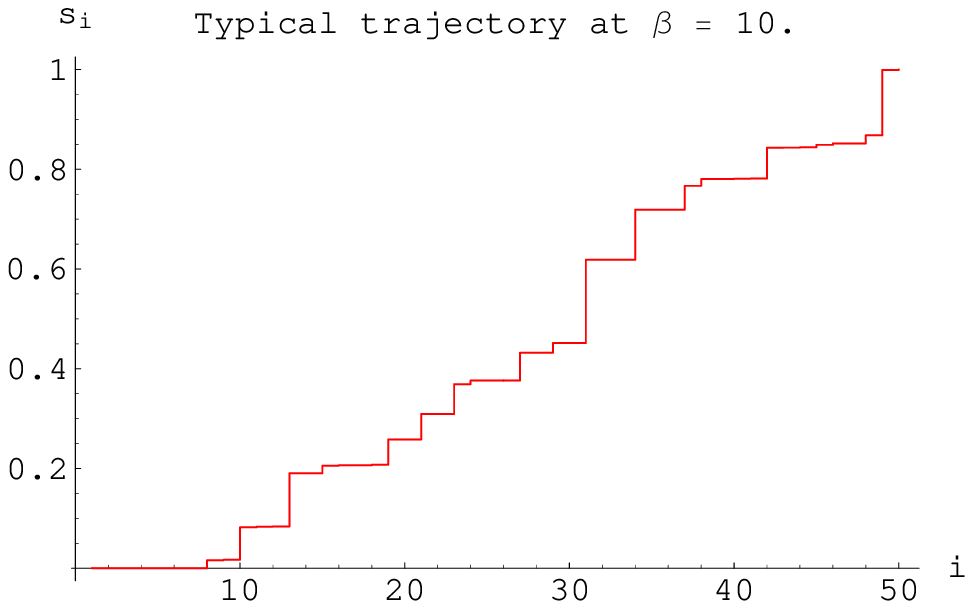}\\
\caption[]{Typical trajectories $s_{i}$ for the measure \rf{measureD} for $\varepsilon = 10^{-6}$ at various $\beta$.}
\label{fi:typicallogbeta_new}
\end{figure}

\begin{figure}
\vspace*{3mm}
\includegraphics[width=5.5cm,angle=-90]{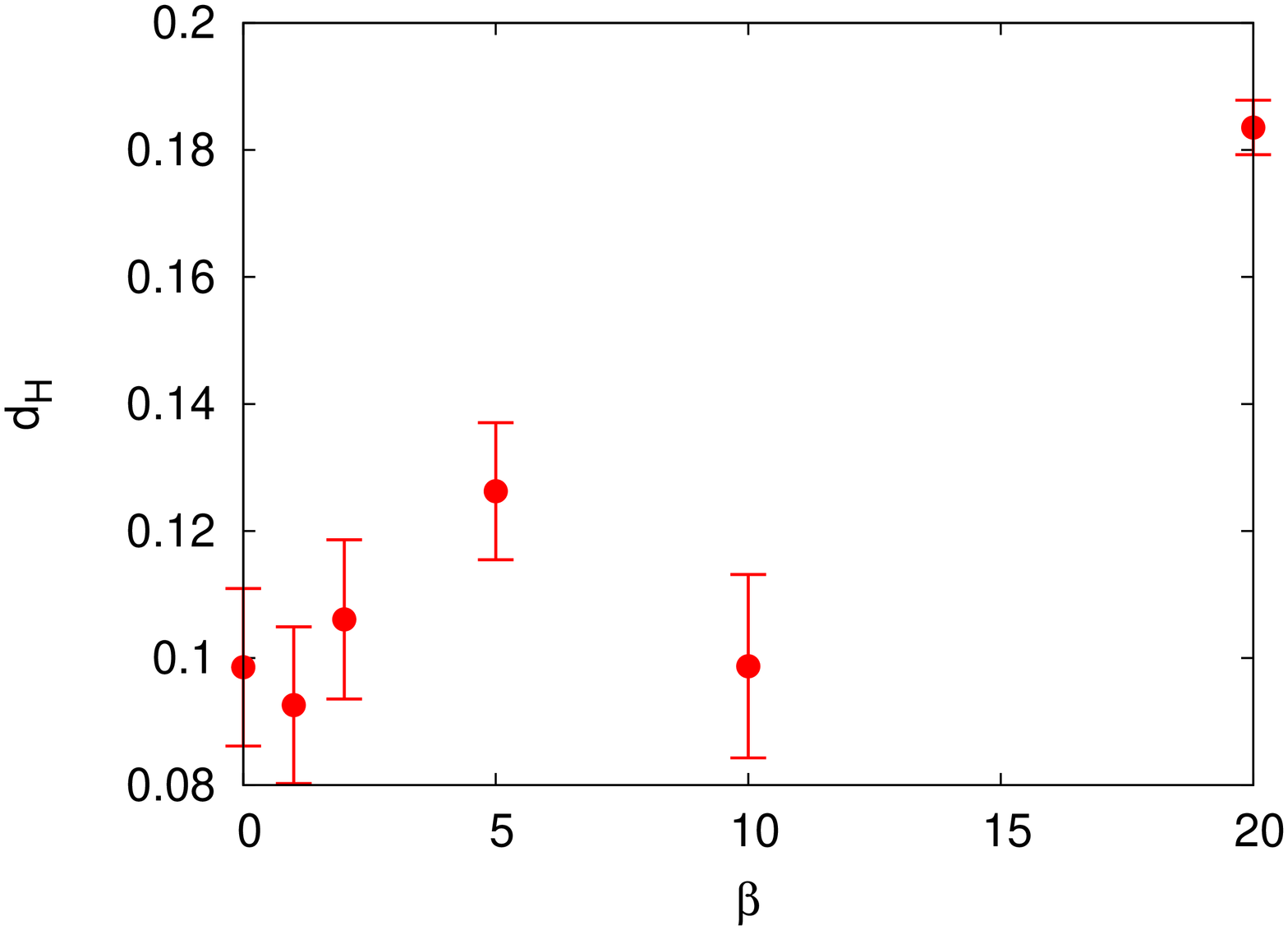}
\hspace*{5mm}
\includegraphics[width=5.5cm,angle=-90]{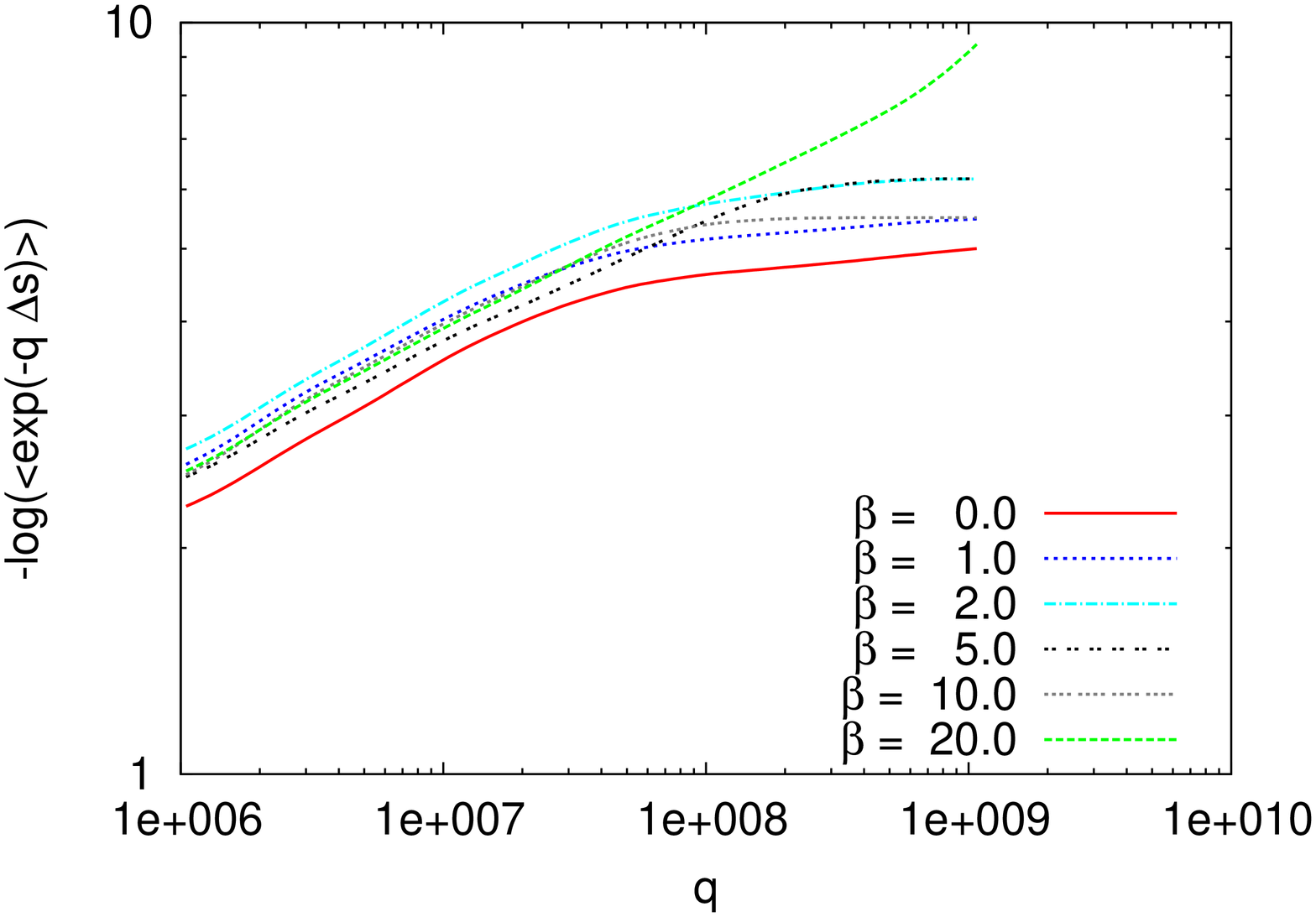}
 \caption[]{Hausdorff dimension of the subordinator with the measure \rf{measureD} versus $\beta$ (left) extracted from the behavior of $-\ln \LA\exp{-q \Delta s}\RA$ at $\varepsilon = 10^{-6}$ (right).}
\label{fi:HDlog-s-beta_new}
\end{figure}

 The average value of \rf{DouglasDiscrNew} is plotted in Fig.~\ref{fi:area_vs_beta_new} versus $\beta$ at $\varepsilon = 10^{-6}$ and $N = 50$. 
Typical trajectories are plotted in Fig.~\ref{fi:typicallogbeta_new}, and their Hausdorff dimensions are shown in Fig.~\ref{fi:HDlog-s-beta_new}. We can see that the mean area decreases with $\beta$, as expected. Typical trajectories also become more smooth, as is indicated by a slow growth of their Hausdorff dimension.

\section{Conclusions}

 Our main result is that the path integral over reparametrizations is tractable pretty much like the usual path integrals with the Wiener measure, while the measure is different. The typical trajectories have, in general, discontinuities of the type of L\'evy's flights and zero Hausdorff dimension, as distinct from the Brownian trajectories whose Hausdorff dimension is equal to two for the usual path integrals.

 We have considered two types of the path integrals over reparametrizations: when the integrand equals one or is given by the exponential of the Douglas integral for a circle. The former emerges in calculations~\cite{MO08,MO09} of QCD scattering amplitudes, while the latter is present in the effective string ansatz~\cite{Pol97} for the asymptotically large Wilson loops. The Hausdorff dimension of typical trajectories for the latter is expected to decrease from one to zero with decreasing radius of the circle.

 Our analysis of the path integral over reparametrizations is based on representing it as a (subordinator) stochastic process, whose probability density has an infinite variance which results in several subtleties (like the central limit theorem or the law of large numbers are not applicable). Nevertheless, the case of the integrand equal to one can be studied analytically and the case of a nontrivial integrand can be studied numerically by applying the Metropolis--Hastings algorithm. The results justify the treatment of the path integral over reparametrizations in Ref.~\cite{MO09}.

 When the exponential of the discretized Douglas functional is included into the measure, it makes the reparametrizations more smooth in some range of circle radii, as should be expected. However, at larger radii the behavior of the average value of the ``naively'' discretized Douglas functional becomes anomalous and decreases below the minimal area $A_{\rm min}[s] = \pi R^{2}$. The reason is that Douglas' theorem is no longer applicable in the discretized case and typical trajectories have jumps which decrease their Hausdorff dimension. This no longer happens, when the discretization of reparametrizing functions is polygon-wise rather than step-wise. Such a situation might be similar to the discretization of the fermionic action in lattice gauge theory, where a ``naively'' discretized action necessarily looses many properties of the continuum theory. 
A slow convergence of the average area to its asymptote may indicate that one should think of a completely different way to integrate over reparametrizations in the effective string ansatz.

 Although our numerical simulations of the effective string ansatz are only for a circle, the method can be straightforwardly applied for other contours. It would be also interesting to apply numerical methods for evaluations of $2\to2$ meson scattering amplitudes in the Regge kinematical regime at finite quark mass, when a spin-dependent kernel is involved~\cite{MO08}, in particular, for extracting the value of the intercept of the quark-antiquark Regge trajectory.

\begin{acknowledgments}
We are indebted to Poul Olesen for sharing his insight and to Alexander Gorsky, Fedor Gubarev, Andrey Mironov, Mikhail Polikarpov, Vladimir Shevchenko, and Konstantin Zarembo for useful discussions. This work was partially supported by the grant for scientific schools No. NSh-679.2008.2 and by the Russian Federal Agency for Nuclear Power. P.B.\ was also supported by the personal grant of the ``Dynasty'' foundation.
\end{acknowledgments}



\end{document}